\documentclass[aps,showpacs,twocolumn,amsmath,amssymb,superscriptaddress,prl,nofootinbib]{revtex4-2}

\usepackage{graphicx}

\usepackage{bm}
\usepackage{color} 
\usepackage{comment}
\usepackage{physics}
\usepackage[T1]{fontenc}
\usepackage{hyperref}

\usepackage{multirow}
\usepackage{pifont} 

\usepackage{lineno}

\begin{document}

\title{Unconventional orbital currents and torques due to ferro-rotational orbital textures}

\date{\today}

\author{Daegeun~Jo}
\affiliation{Department of Physics and Astronomy, Uppsala University, P.O. Box 516, SE-75120 Uppsala, Sweden}
\affiliation{Wallenberg Initiative Materials Science for Sustainability, Uppsala University, SE-75120 Uppsala, Sweden}

\author{Peter~M.~Oppeneer}
\affiliation{Department of Physics and Astronomy, Uppsala University, P.O. Box 516, SE-75120 Uppsala, Sweden}
\affiliation{Wallenberg Initiative Materials Science for Sustainability, Uppsala University, SE-75120 Uppsala, Sweden}


\begin{abstract}
Orbital angular momentum transport has emerged as a promising route for manipulating magnetic devices, yet its generation has largely relied on the conventional orbital Hall effect. Here, we show that ferro-rotational order enables the electrical generation of unconventional orbital currents. These orbital currents represent the orbital counterparts of spin currents due to ferromagnetic order, but arise from rotation-induced symmetry breaking rather than time-reversal symmetry breaking or spin-orbit coupling. Using tight-binding models, we identify the underlying intrinsic, nonrelativistic mechanism categorized as
an electric hexadecapole moment and corroborate our findings with first-principles calculations for the ferro-rotational material TiAu$_4$. We further show that these rotation-induced orbital currents lead to surface orbital accumulation and unconventional orbital torque in a ferro-rotational/ferromagnetic metallic bilayer, allowing deterministic field-free switching. Our findings unveil a novel pathway for generating orbital currents beyond the conventional orbital Hall effect, broadening the landscape of orbitronics research to include novel ferroic materials and higher-order electric multipoles.
\end{abstract}

\maketitle

\let\thefootnote\relax\footnotetext{Corresponding authors: Daegeun Jo (daegeun.jo@physics.uu.se) and Peter~M.~Oppeneer (peter.oppeneer@physics.uu.se)}
	
\section{Introduction}
\begin{figure*}[bt!]
	\center\includegraphics[width=1\textwidth]{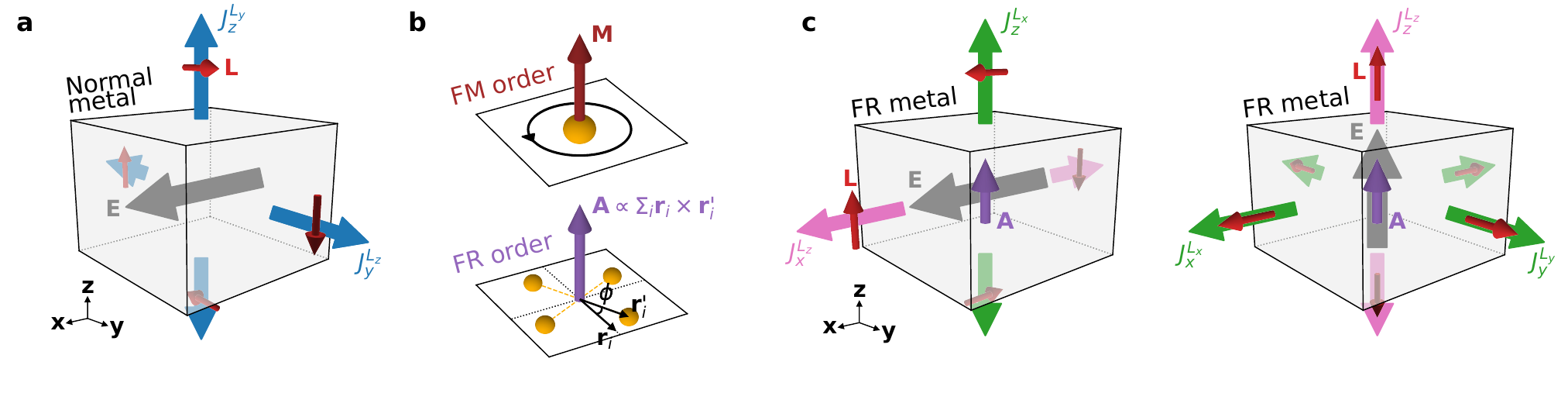}
	\caption{Schematic of rotation-induced orbital currents in FR systems. \textbf{a} Conventional orbital Hall effect in a normal metal. The electric field ($\mathbf{E}$), orbital current ($\mathbf{J}^\mathbf{L}$), and OAM ($\mathbf{L}$) are mutually orthogonal. \textbf{b} FM (top) and FR (bottom) orders. The FM order is described by magnetization $\mathbf{M}$, which is odd under $\mathcal{T}$. The FR order, arising from a rotational displacement of atoms by an angle $\phi$ within the unit cell, is characterized by an axial vector $\mathbf{A}$ along the rotational axis, which is even under $\mathcal{T}$. \textbf{c} Rotation-induced orbital currents in a FR metal for $\mathbf{E} \perp \mathbf{A}$ (left) and $\mathbf{E} \parallel \mathbf{A}$ (right). Pink and green arrows indicate longitudinal and unconventional Hall (transverse) components, respectively.
	}
	\label{fig1} 
\end{figure*}

Over the past few decades, the transport of electron angular momentum has emerged as a central theme in both condensed matter physics and information technology. Spin currents, i.e., flows of spin angular momentum, have become a fundamental building block of spintronics research, enabling efficient manipulation of magnetization and advancing prospects for low-power memory devices~\cite{manchon2019current, hirohata2020review}. Electrical generation of spin currents typically requires either magnetic order---breaking time-reversal ($\mathcal{T}$) symmetry---or relativistic spin-orbit coupling (SOC). By contrast, the orbital Hall effect (OHE) (Fig.~\ref{fig1}a) has been predicted~\cite{bernevig2005orbitronics, kontani2009giant, go2018intrinsic, jo2018gigantic, bhowal2020intrinsic, canonico2020orbital, bhowal2021orbital, salemi2022first} and observed~\cite{choi2023observation, lyalin2023magneto, idrobo2024direct} to generate orbital currents even in nonmagnetic materials without relying on SOC. Because of its nonrelativistic origin, light elements can be utilized as efficient sources of orbital currents, opening new avenues for magnetic device applications via orbital torque~\cite{go2020orbital, kim2021nontrivial, ding2020harnessing, lee2021orbital, lee2021efficient, sala2022giant, hayashi2023observation} and thereby directing growing attention to orbital angular momentum (OAM) transport~\cite{choi2023observation, lyalin2023magneto, seifert2023time, hayashi2024observation, idrobo2024direct}.

Despite rapid progress in orbitronics research~\cite{go2021orbitronics, jo2024spintronics, atencia2024orbital}, most studies of electrically generated orbital currents remain confined to the conventional OHE, where the electric field, current flow, and angular momentum polarization are mutually orthogonal (Fig.~\ref{fig1}a). By contrast, spin currents with nonorthogonal configurations---referred to here as ``unconventional''---have become extensively explored for both their symmetry and device implications~\cite{wimmer2015spin, seemann2015symmetry, amin2016b, macneill2017control, humphries2017observation, zelezny2017spin, baek2018spin, amin2018interface, kimata2019magnetic, hernandez2021efficient, bose2022tilted, roy2022unconventional, hayami2022electric, hayami2023unconventional}. In particular, ferromagnetic (FM) order, which breaks $\mathcal{T}$ symmetry, serves as a controllable source of magnetization-induced spin currents~\cite{mook2020origin,  Salemi2021, salemi2022theory, park2022spin}. While analogous orbital currents can coexist with these spin currents due to SOC~\cite{Salemi2021,salemi2022theory}, a \emph{purely nonrelativistic} mechanism for generating orbital currents beyond conventional Hall components has yet to be identified.

Here we propose ferro-rotational (FR) order~\cite{gopalan2011rotation, johnson2011cu, johnson2012giant, hlinka2014eight, cheong2018broken} (Fig.~\ref{fig1}b)---also known as ferro-axial order---as a new pathway for orbital current generation, in analogy to magnetization-induced spin currents in FM materials. FR order has drawn much interest recently \cite{hayami2022electric,hayami2023unconventional,jin2020observation, hayashida2020visualization, singh2025ferroaxial, du2026electric}.
It emerges from a static structural rotation and naturally couples to the orbital degree of freedom. A key observation is the symmetry correspondence between FR and FM orders: both break vertical mirror planes while preserving the horizontal mirror plane and inversion ($\mathcal{P}$) symmetry. From this relation, together with $\mathcal{T}$-even and $\mathcal{T}$-odd parities of FR and FM orders, respectively, one can directly infer that $\mathcal{T}$-even angular momentum currents in FR systems can emerge as natural counterparts of the $\mathcal{T}$-odd currents in FM systems. Indeed, unconventional $\mathcal{T}$-even spin currents have been predicted in certain nonmagnetic systems~\cite{wimmer2015spin, seemann2015symmetry, roy2022unconventional}, recently attributed to the FR order combined with SOC~\cite{hayami2022electric, hayami2023unconventional}. However, a crucial 
distinction is that the FR order can couple to the electron orbital without relying on SOC, thereby enabling what we term \emph{rotation-induced orbital currents}, constituting a purely nonrelativistic mechanism. 

In this work, we demonstrate the electrical generation of  unconventional orbital currents in FR systems that preserve both $\mathcal{P}$ and $\mathcal{T}$ symmetries. Symmetry arguments reveal that these $\mathcal{T}$-even rotation-induced orbital currents---analogous to $\mathcal{T}$-odd magnetization-induced spin currents in FM systems---manifest as (i) longitudinal orbital currents polarized along the FR axis and (ii) unconventional orbital Hall currents with polarization collinear with either the charge or orbital current [e.g., see (Fig.~\ref{fig1}c). Tight-binding calculations show that these effects are driven by an electric hexadecapole (16-pole) moment arising from the FR order, through an intrinsic and nonrelativistic mechanism. To corroborate these findings, we perform first-principles calculations for the FR material TiAu$_4$. Finally, we explore the experimental implications of rotation-induced orbital currents by studying a FR/FM bilayer within a tight-binding framework, demonstrating current-induced orbital accumulation in the FR layer as well as orbital torque in the FM layer. We further suggest that this unconventional orbital torque can enable deterministic, field-free switching of the FM order, pointing to a promising route for orbitronics research based on novel ferroic orders and higher-order electric multipoles.

\section{Results}
\subsection{Symmetry arguments}
In the linear-response regime, the orbital current $\mathbf{J}^{\mathbf{L}}$ (or spin current $\mathbf{J}^{\mathbf{S}}$) generated by an electric field $\mathbf{E}$ is expressed as $J_\alpha^{X_\gamma} = \sigma_{\alpha\beta}^{X_\gamma} E_\beta$, where $X=L$ or $S$. Here, $\alpha$ and $\gamma$ denote the orbital (spin) current flow and polarization directions, respectively. The rank-3 orbital (spin) conductivity tensor $\bm{\sigma}^\mathbf{L}$ ($\bm{\sigma}^\mathbf{S}$) can generally be decomposed into $\mathcal{T}$-even and $\mathcal{T}$-odd contributions, with their nonzero components dictated by the system's symmetry. For example, in a nonmagnetic cubic system with point group $O_h$, only the $\mathcal{T}$-even conventional Hall components of $\sigma_{\alpha\beta}^{X_\gamma}$, where $\alpha$, $\beta$, and $\gamma$ are mutually orthogonal, are symmetrically allowed.

Symmetry breaking due to ferroic orders can induce additional nonzero components of $\bm{\sigma}^\mathbf{X}$. Here, we focus on ferroic orders that preserve $\mathcal{P}$ symmetry, classified into two types~\cite{hlinka2014eight, cheong2018broken}: $\mathcal{T}$-odd FM order and $\mathcal{T}$-even FR order. In a cubic system, the FR and FM orders aligned along the $z$ direction reduce the symmetry, leading to the point group $4/m$ and the magnetic point group $4/mm^\prime m^\prime$, respectively. For both cases, the nonzero components of the total $\bm{\sigma}^\mathbf{X}$ are given by~\cite{wimmer2015spin, seemann2015symmetry, roy2022unconventional, salemi2022theory}:
\begin{gather}
	\bm{\sigma}^{X_x} =
	\begin{pmatrix}
		0 & 0 & \sigma_{xz}^{X_x} \\
		0 & 0 & \sigma_{yz}^{X_x} \\
		\sigma_{zx}^{X_x} & \sigma_{zy}^{X_x} & 0 
	\end{pmatrix},
	\,
	\bm{\sigma}^{X_y} =
	\begin{pmatrix}
		0 & 0 & \sigma_{xz}^{X_y} \\
		0 & 0 & \sigma_{yz}^{X_y} \\
		\sigma_{zx}^{X_y} & \sigma_{zy}^{X_y} & 0 
	\end{pmatrix}, \nonumber \\
	\bm{\sigma}^{X_z} =
	\begin{pmatrix}
		\sigma_{xx}^{X_z} & \sigma_{xy}^{X_z} & 0 \\
		\sigma_{yx}^{X_z} & \sigma_{yy}^{X_z} & 0 \\
		0 & 0 & \sigma_{zz}^{X_z}
	\end{pmatrix}. \label{eq:cond_tensor}
\end{gather}
In addition to the $\mathcal{T}$-even conventional Hall components ($\sigma_{yz}^{X_x} = - \sigma_{xz}^{X_y}$, $\sigma_{zy}^{X_x} = -\sigma_{zx}^{X_y}$, and $\sigma_{xy}^{X_z} = - \sigma_{yx}^{X_z}$), the components induced by ferroic orders can be categorized into two groups: (i) diagonal components ($\sigma_{xx}^{X_z} = \sigma_{yy}^{X_z}$ and $\sigma_{zz}^{X_z}$), describing longitudinal currents polarized along the order parameter (pink arrows in Fig.~\ref{fig1}c), and (ii) off-diagonal components ($\sigma_{xz}^{X_x} = \sigma_{yz}^{X_y}$ and $\sigma_{zx}^{X_x} = \sigma_{zy}^{X_y}$), representing unconventional Hall currents, where the polarization is collinear with either $ \mathbf{E}$ or $ \mathbf{J}^\mathbf{X}$ (green arrows in Fig.~\ref{fig1}c). It is worth noting that, in the presence of the first-type longitudinal components $\sigma_{\alpha\alpha}^{X_\beta}$, the second-type Hall components take the form $ \sigma_{\beta\alpha}^{X_\alpha} - \delta_{\alpha\beta} \sigma_{\gamma\alpha}^{X_\gamma} $. This implies a conversion of a primary current  $J_\alpha^{X_\beta}$ into a secondary current $J_\beta^{X_\alpha}$ (or of $J_\alpha^{X_\alpha}$ into $J_\beta^{X_\beta}$) for $\alpha \neq \beta$, corresponding to spin swapping~\cite{lifshits2009swapping} or orbital swapping~\cite{ning2025orbital}.

These ferroic-order-induced currents inherit the $\mathcal{T}$-parity of the associated order parameters. In $\mathcal{T}$-odd FM metals, $\mathcal{T}$-odd longitudinal spin currents are electrically generated due to the nonrelativistic spin-polarized band structure. Additionally, $\mathcal{T}$-odd unconventional spin Hall currents---also known as the magnetic spin Hall effect~\cite{mook2020origin, Salemi2021, salemi2022theory} or spin swapping~\cite{lifshits2009swapping}---arise from SOC. These magnetization-induced spin currents in FM metals can accompany the relativistic $\mathcal{T}$-odd orbital currents via SOC, e.g., the magnetic OHE~\cite{Salemi2021, salemi2022theory}. 
In contrast, in $\mathcal{T}$-even FR systems, the rotation-induced orbital currents can be generated, including the longitudinal orbital currents and unconventional orbital Hall (or orbital swapping) currents. Importantly and distinctively, they require neither broken $\mathcal{T}$ nor SOC, as will be demonstrated.

\subsection{Rotation-induced longitudinal orbital current}
\begin{figure}[t]
	\center\includegraphics[width=0.5\textwidth]{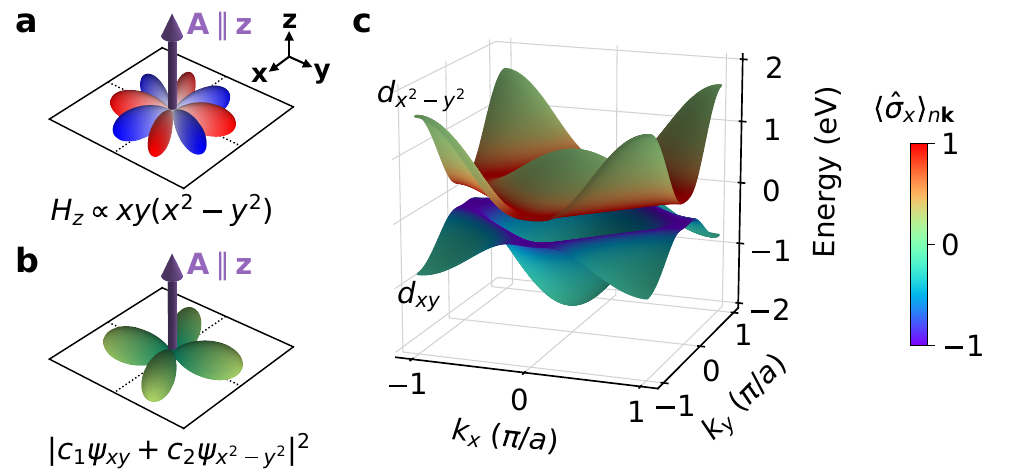}
	\caption{Electric hexadecapole moment in a FR system. Illustrations of \textbf{a} electric hexadecapole moment $H_z$ and \textbf{b} wave function of a rotated $d_{xy}$ or $d_{x^2-y^2}$ orbital in the presence of the FR order parameter $\mathbf{A}$. \textbf{c} Band structure of the two-orbital model described by Eq.~\eqref{eq:H_toy}, with color indicating the expectation value of $\hat{\sigma}_x$.
	}
	\label{fig2} 
\end{figure}

To see how the orbital current can be generated in FR systems, we introduce a minimal tight-binding model with a relevant order parameter. The OAM dynamics can be driven by multipole degrees of freedom~\cite{han2022orbital, han2025harnessing}. Although the FR order is often described by an axial vector, such as the electric toroidal moment~\cite{hlinka2014eight, cheong2018broken, jin2020observation, hayashida2020visualization}, we focus here on another emergent multipole in FR systems: the electric hexadecapole moment (rank-4)~\cite{hayami2022electric, hayami2023unconventional}, $H_z \propto xy(x^2-y^2)$ (Fig.~\ref{fig2}a), which is even under $\mathcal{P}$ and $\mathcal{T}$. The quantum mechanical operator for this can be constructed by replacing $\mathbf{r} = (x,y,z)$ with the OAM operators $\hat{\mathbf{L}} = (\hat{L}_x, \hat{L}_y, \hat{L}_z)$~\cite{kusunose2008description}. Accordingly, we define an atomic-site electric hexadecapole moment operator as
\begin{equation}
\hat{H}_z \equiv \frac{1}{12\hbar^4} \{ \{\hat{L}_x , \hat{L}_y \} , \hat{L}_x^2 - \hat{L}_y^2  \}, \label{eq:hexadecapole}
\end{equation}
where $\{\hat{a},\hat{b}\} = \hat{a}\hat{b} + \hat{b}\hat{a}$ and $\hbar$ is the reduced Planck constant. Note that $\hat{H}_z$ can emerge under the point group $4/m$ exhibiting the FR order along the $z$ direction~\cite{hayami2022electric, hayami2023unconventional}. In the atomic $d$-orbital basis \{$d_{xy}, d_{yz}, d_{zx}, d_{x^2-y^2}, d_{3z^2 - r^2}$\}, Eq.~\eqref{eq:hexadecapole} simplifies to $\hat{H}_z^{d} \equiv \vert d_{xy} \rangle \langle d_{x^2-y^2 }\vert + \vert d_{x^2-y^2} \rangle \langle d_{xy} \vert$, which implies that $\hat{H}_z$ hybridizes orbital wave functions, effectively rotating them around the $z$-axis, as illustrated in Fig.~\ref{fig2}b.

Let us introduce a two-dimensional square lattice tight-binding model incorporating $\hat{H}_z^d$. We adopt a minimal two-orbital basis \{$\vert d_{xy} \rangle , \vert d_{x^2-y^2} \rangle$\} for describing $\hat{H}_z^d$. Considering only nearest-neighbor hopping, the Hamiltonian is given by
\begin{equation} \label{eq:H_toy}
	\hat{\mathcal{H}}(\mathbf{k}) = (\cos k_x a + \cos k_y a) ( t_+ \hat{\sigma}_0 + t_- \hat{\sigma}_z ) + \Delta \hat{\sigma}_x,
\end{equation}
where $\mathbf{k}$ is the crystal momentum, $a$ is the lattice constant, $\hat{\sigma}_0$ is the identity matrix, $\hat{\bm{\sigma}} = (\hat{\sigma}_x, \hat{\sigma}_y, \hat{\sigma}_z)$ are the pseudospin Pauli matrices, and $t_\pm \equiv t_\pi^d \pm (3t_\sigma^d + t_\delta^d)/4 $ is determined by the Slater-Koster hopping parameters $(t_\sigma^d, t_\pi^d, t_\delta^d) = (-0.5, 0.2, -0.1)$ in units of eV. The $\hat{\sigma}_z$ term in Eq.~\eqref{eq:H_toy} accounts for the crystal field that splits $d_{xy}$ and $d_{x^2-y^2}$ levels. The effect of the FR order along the $z$ direction is incorporated through $\hat{\sigma}_x$, which is equivalent to $\hat{H}_z^d$ in the two-orbital basis, with its magnitude set by $\Delta = 0.1$~eV.  Figure~\ref{fig2}c shows that a gap between $d_{xy}$ and $d_{x^2-y^2}$ bands is opened due to the electric hexadecapole moment. Near the gap, the eigenstates $\ket{\psi_{\pm,\mathbf{k}}} \approx (\ket{d_{xy}} \pm \ket{d_{x^2-y^2}} )/\sqrt{2} $, with energies $\epsilon_{\pm ,\mathbf{k}} \approx \pm \Delta$, yield the expectation values $ \langle \hat{\bm{\sigma}} \rangle_{\pm ,\mathbf{k}} \equiv \langle \psi_{\pm ,\mathbf{k}}  \vert \hat{\bm{\sigma}} \vert \psi_{\pm ,\mathbf{k}}  \rangle   \approx (\pm 1, 0,0)$. We note that $\hat{L}_z^\mathrm{sub} \equiv -2\hbar \hat{\sigma}_y $ corresponds to the submatrix of $\hat{L}_z$ that is defined in the full $d$-orbital basis, so $ \hat{\sigma}_y$ effectively captures the out-of-plane OAM.

Here, we derive an intuitive picture of how an electric field $\mathbf{E}$ drives the dynamics of $\hat{\bm{\sigma}}$ for a single Bloch state near the gap. Under $\mathbf{E}=E \hat{\mathbf{x}}$, an electron with charge $-e$ after time $\delta t$ acquires momentum $\delta k_x = -eE \delta t / \hbar$, leading to the perturbation $ \delta \hat{\mathcal{H}} = - \delta k_x a  \sin k_x a (t_+ \hat{\sigma}_0 + t_- \hat{\sigma}_z)$. The dynamics of $\hat{\bm{\sigma}}$ follows the Bloch equation $d \langle \hat{\bm{\sigma}} \rangle_{\pm, \mathbf{k}} /dt = (2/\hbar) (\mathbf{B}(\mathbf{k}) \times \langle \hat{\bm{\sigma}} \rangle_{\pm, \mathbf{k}} ) $, where $\mathbf{B}(\mathbf{k}) $ is the effective magnetic field satisfying $\hat{\mathcal{H}} + \delta \hat{\mathcal{H}} = \mathbf{B} \cdot \hat{\bm{\sigma}}$, with $dB_z/dt = (eEat_- /\hbar) \sin k_x a $ arising from the electric-field-induced crystal field variation. In the vicinity of the band gap, with an initial condition $\langle \hat{\bm{\sigma}} \rangle_{\pm,\mathbf{k}} = (\pm1,0,0) $, the solutions for small deviations from equilibrium are given by $\langle \hat{\sigma}_x \rangle_{\pm, \mathbf{k}} \approx \pm 1$, $\langle \hat{\sigma}_z \rangle_{\pm, \mathbf{k}} \approx B_z(\mathbf{k}) / \Delta$, and 
\begin{equation}\label{eq:neq_sigma_y}
\langle \hat{\sigma}_y \rangle_{\pm, \mathbf{k}} = \pm \frac{\hbar}{2\Delta^2}\frac{dB_z(\mathbf{k})}{dt} \approx \pm \frac{eE a t_-}{2 \Delta^2} \sin k_x a.
\end{equation}
This result shows that the electric hexadecapole moment undergoes precession due to the intrinsic crystal field that acts as a current-induced effective field, generating the nonequilibrium OAM $\langle \hat{L}_z^\mathrm{sub} \rangle_{\pm, \mathbf{k}} = -2\hbar \langle \hat{\sigma}_y \rangle_{\pm, \mathbf{k}}$. This behavior resembles spin dynamics in FM systems under an intrinsic spin-orbit field~\cite{kurebayashi2014antidamping}, although the effect here is nonrelativistic. Note that $\langle \hat{\sigma}_y \rangle_{\pm, \mathbf{k}}$ in Eq.~\eqref{eq:neq_sigma_y} diverges as $\Delta \rightarrow 0$, but the net value vanishes as the gap closes. 

Although the net OAM (or $\hat{\sigma}_y $) vanishes upon $\mathbf{k}$-integration, the net orbital current remains finite, leading to a nonzero $\sigma_{xx}^{L_z}$. The conventional orbital current operator is defined as $ \hat{\mathbf{J}}^{L_\gamma} \equiv \frac{1}{2} \{ \hat{\mathbf{v}}, \hat{L}_\gamma \} $, where $\hat{\mathbf{v}}$ is the velocity operator. Substituting $\hat{L}_z^\mathrm{sub}$ and $\hat{\mathbf{v}} = (1/\hbar) \partial_\mathbf{k} \hat{\mathcal{H}}(\mathbf{k}) $, the longitudinal orbital current to first order in $E$ is given by
\begin{equation}\label{eq:neq_orbital_current}
	\langle \hat{J}_x^{L_z} \rangle_{\pm, \mathbf{k}} =  \pm (\frac{e E a^2 t_+ t_-}{ \Delta^2} )\sin^2 k_x a ,
\end{equation}
Integration of Eq.~\eqref{eq:neq_orbital_current} over $\mathbf{k}$-space yields a finite value, confirming the emergence of a rotation-induced orbital current driven by an intrinsic, nonrelativistic mechanism associated with a higher-order electric multipole. 

\subsection{Unconventional orbital Hall current}
\begin{figure}[t]
	\center\includegraphics[width=0.5\textwidth]{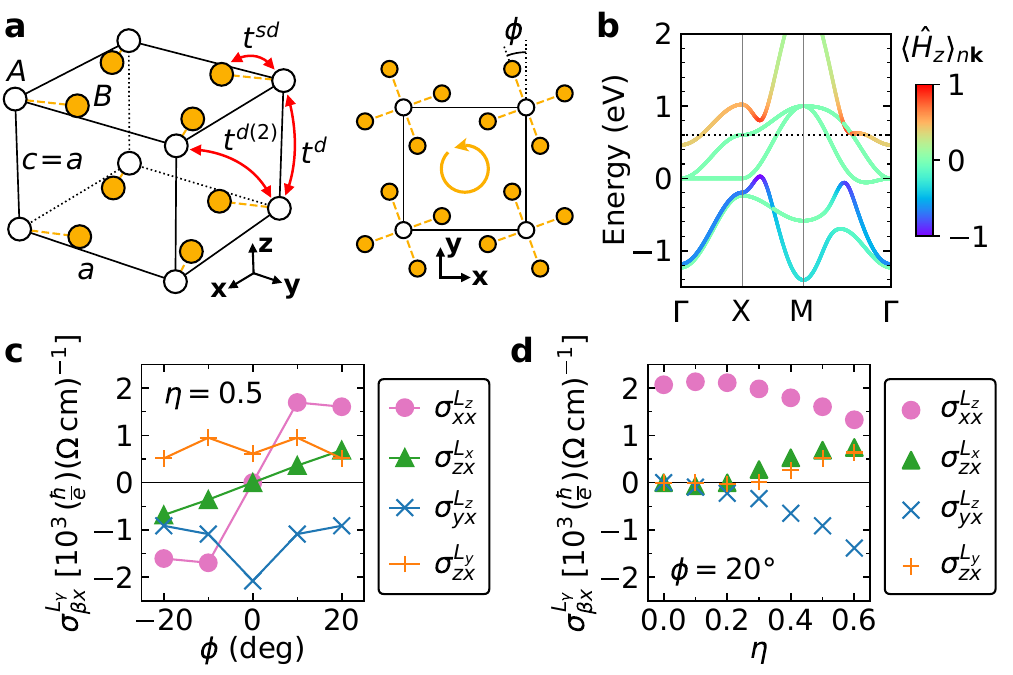}
	\caption{Three-dimensional FR model. \textbf{a} Crystal structure of the tight-binding model. Red arrows depict the hopping pairs considered in the model. The right panel illustrates the $xy$-plane structure, exhibiting the FR order along $\hat{\mathbf{z}}$. \textbf{b} Band structure of the $d$-bands, with color representing the expectation value of $\hat{H}_z$. The dotted line indicates the assumed chemical potential of 0.6 eV. \textbf{c, d} Orbital conductivities $\sigma_{\beta x}^{L_\gamma}$ for different values of \textbf{c} $\phi$ and \textbf{d} $\eta$.
	}
	\label{fig3} 
\end{figure}

Additional orbital currents can emerge in multi-orbital systems exhibiting richer orbital texture. To explore this, we construct a three-dimensional tight-binding model for a FR system with the point group $4/m$ (Fig.~\ref{fig3}a; see Methods and Supplementary Note 1), which constrains $\bm{\sigma}^\mathbf{L}$ as given in Eq.~\eqref{eq:cond_tensor}. The tetragonal unit cell consists of $A$ atoms with five $d$ orbitals, and $B$ atoms with an $s$ orbital. The FR order along the $z$-axis arises from a rotational displacement of the four $B$ atoms by an angle $\phi$. The hopping pairs included in the model are shown in Fig.~\ref{fig3}a. The next-nearest-neighbor hopping between $d$ orbitals gives rise to the momentum-dependent $d$-orbital texture responsible for the conventional OHE~\cite{go2018intrinsic, jo2018gigantic}. Its amplitude is assumed proportional to that of the nearest-neighbor hopping, with the proportionality factor $\eta$ initially set to 0.5. The hopping between $s$ and $d$ orbitals, which depends on $\phi$, characterizes the FR order.

Figure~\ref{fig3}b shows the band structure of this model with $\phi=20^\circ$, which exhibits a nonzero expectation value of $\hat{H}_z$ [defined in Eq.~\eqref{eq:hexadecapole}] in equilibrium. Unlike earlier works~\cite{hayami2022electric, hayami2023unconventional}, where $\hat{H}_z$ was manually introduced into the Hamiltonian, in this model, it naturally emerges from structural rotation. It is noteworthy that downfolding our Hamiltonian into the two-dimensional $d$-orbital subspace yields a term proportional to $\phi \hat{H}_z$ for small $\phi$ (see Supplementary Note 2), revealing a direct connection between the electric hexadecapole moment and the FR order.

We now proceed to compute the $\mathcal{T}$-even part of the orbital conductivity tensor $\bm{\sigma}^\mathbf{L}$ using the Kubo formula~\cite{go2018intrinsic} (see Methods). Figure~\ref{fig3}c presents numerical results for the nonzero orbital conductivity components $\sigma_{\beta x}^{L_\gamma}$ for different values of $\phi$, with $\mathbf{E} \parallel \hat{\mathbf{x}}$. The longitudinal ($\sigma_{xx}^{L_z}$) and unconventional orbital Hall ($\sigma_{zx}^{L_x}$) components  (e.g., see Fig.~\ref{fig1}c), represented by pink circles and green triangles, respectively, vanish at $\phi=0$ and reverse sign under FR-order-reversal ($\phi \rightarrow -\phi$). In contrast, the conventional orbital Hall components, indicated by blue $\times$ and orange $+$ symbols, remain finite at $\phi=0$ and are invariant under FR-order-reversal. These results clearly demonstrate that rotation-induced OHE and conventional OHE have distinct physical origins, while both are $\mathcal{T}$-even and nonrelativistic. 

To further investigate the mechanism behind rotation-induced orbital currents, we compute $\sigma_{\beta x}^{L_\gamma}$ for different values of $\eta$, which controls the next-nearest-neighbor hopping amplitudes, while fixing $\phi = 20^\circ$ (Fig.~\ref{fig3}d).  We find that only the longitudinal component remains finite for $\eta = 0$, indicating that it arises solely from the FR order, specifically the electric hexadecapole moment, as demonstrated by our two-orbital model. On the other hand, both conventional and unconventional Hall components emerge as $\eta$ increases, suggesting that the rotation-induced OHE requires not only the FR order but also the orbital texture responsible for the conventional OHE. This phenomenon can be understood in terms of nonrelativistic orbital swapping~\cite{ning2025orbital}---an orbital analog of spin swapping~\cite{lifshits2009swapping}. It has been shown that in FM metals, a spin-polarized current $J_{x}^{S_z}$ is converted into a swapped spin current $J_{z}^{S_x}$ through the interplay of the orbital texture and SOC~\cite{park2022spin}. Similarly, our results show that the longitudinal orbital current $J_x^{L_z}$, induced by the FR order, is converted into the unconventional orbital Hall current $J_z^{L_x}$ (or $J_z^{L_z}$ into $J_x^{L_x}$ when $\mathbf{E} \parallel \hat{\mathbf{z}}$) via the orbital texture. Notably, this conversion does not require SOC, in contrast to spin swapping. 

\begin{figure}[t]
	\center\includegraphics[width=0.5\textwidth]{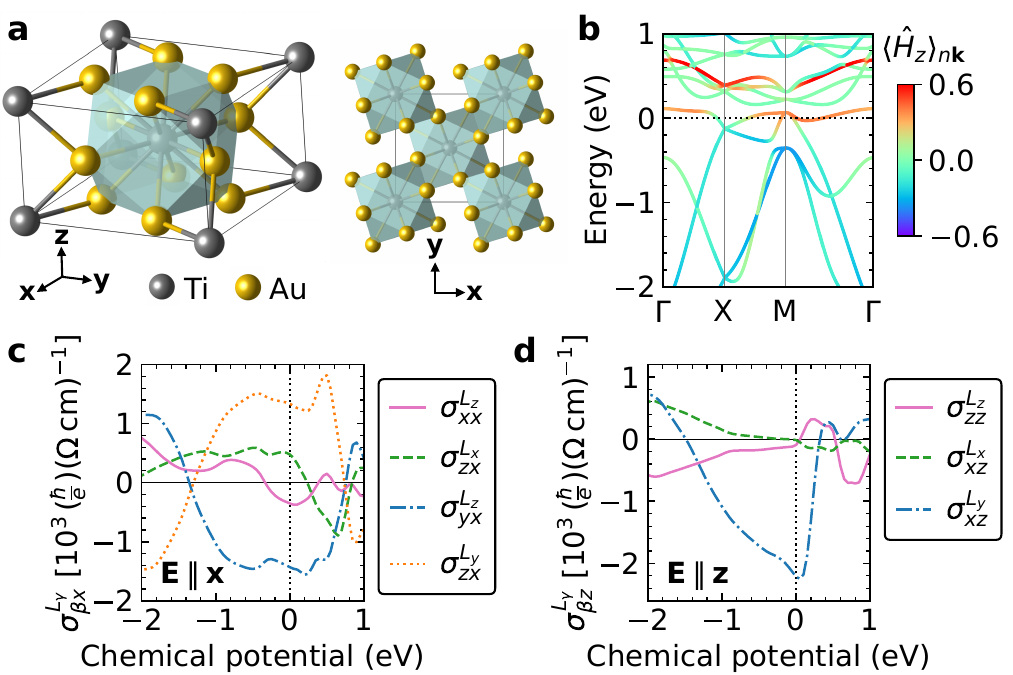}
	\caption{First-principles calculations for TiAu$_4$. \textbf{a} Crystal structure of tetragonal TiAu$_4$. \textbf{b} Band structure near the Fermi level (0 eV), with color representing the expectation value of $\hat{H}_z$, as defined in Eq.~\eqref{eq:hexadecapole}. \textbf{c, d} Nonzero orbital conductivity components $\sigma_{\beta \alpha}^{L_\gamma}$ as functions of the chemical potential, for an electric field $\mathbf{E}$ applied along \textbf{c} the $x$ direction and \textbf{d} the $z$ direction.}
	\label{fig4} 
\end{figure}

\subsection{First-principles calculation for TiAu$_4$}
Next, we investigate the FR material candidate,
tetragonal TiAu$_4$ (space group $I4/m$)~\cite{murray1983au} using first-principles calculations (see Methods). The crystal structure exhibits the FR order along the $z$-axis (Fig.~\ref{fig4}a), leading to a nonzero electric hexadecapole moment (Fig.~\ref{fig4}b). The orbital conductivity tensor $\bm{\sigma}^\mathbf{L}$ takes the same form as Eq.~\eqref{eq:cond_tensor}, with seven independent nonzero components of $\sigma_{\beta \alpha}^{L_\gamma}$, including those for $\alpha=x$ ($\mathbf{E} \parallel \hat{\mathbf{x}}$) and $\alpha = z$ ($\mathbf{E} \parallel \hat{\mathbf{z}}$). The rotation-induced orbital currents associated with these components are illustrated in the left and right panels in Fig.~\ref{fig1}c, respectively. The components for $\mathbf{E} \parallel \hat{\mathbf{y}}$ are related to those for $\mathbf{E} \parallel \hat{\mathbf{x}}$ by four-fold rotational symmetry about the $z$-axis. By evaluating the Kubo formula, the nonzero components $\sigma_{\beta\alpha}^{L_\gamma}$ are obtained as functions of the chemical potential.  For $\mathbf{E} \parallel \hat{\mathbf{x}}$ (Fig.~\ref{fig4}c), the conventional Hall components exceed $1000 \ (\hbar/e) (\Omega \, \mathrm{cm})^{-1}$ at the Fermi level. Additionally, we identify rotation-induced components, including the longitudinal orbital conductivity $\sigma_{xx}^{L_z} = -350 \ (\hbar/e) (\Omega \, \mathrm{cm})^{-1}$ and the unconventional orbital Hall conductivity $\sigma_{zx}^{L_x} = 480 \ (\hbar/e) (\Omega \, \mathrm{cm})^{-1}$. For $\mathbf{E} \parallel \hat{\mathbf{z}}$ (Fig.~\ref{fig4}d), the rotation-induced components are smaller, with $\sigma_{zz}^{L_z} = -90 \ (\hbar/e) (\Omega \, \mathrm{cm})^{-1}$ and $\sigma_{xz}^{L_x} = -10 \ (\hbar/e) (\Omega \, \mathrm{cm})^{-1}$. The magnitude of the unconventional terms depends on the FR orbital texture (e.g., see Fig.~\ref{fig3}), motivating further materials exploration.

While the orbital conductivity is fully nonrelativistic, the corresponding nonzero components of the spin conductivity tensor can manifest, too, due to SOC. When SOC is present, not only the electric multipole moments but also the atomic-site electric toroidal moments, defined in the spinful basis, can emerge from the FR order, contributing to the $\mathcal{T}$-even spin current generation~\cite{hayami2022electric, hayami2023unconventional}. A key distinction, however, is that the spin conductivity vanishes in the absence of SOC, whereas the orbital conductivity remains largely unaffected by SOC due to its nonrelativistic origin (see Supplementary Note 3). Furthermore, the $\mathcal{T}$-even orbital conductivity arises purely from the interband contribution, which is robust against scattering time (see Supplementary Note 3), while possible extrinsic contributions~\cite{tang2024role} are not considered. By contrast, the $\mathcal{T}$-odd conductivity in FM systems is dominated by the intraband contribution that scales with the scattering time~\cite{salemi2022theory}, but it is prohibited here by $\mathcal{T}$ invariance.

\begin{figure*}[hbt!]
	\center\includegraphics[width=1\textwidth]{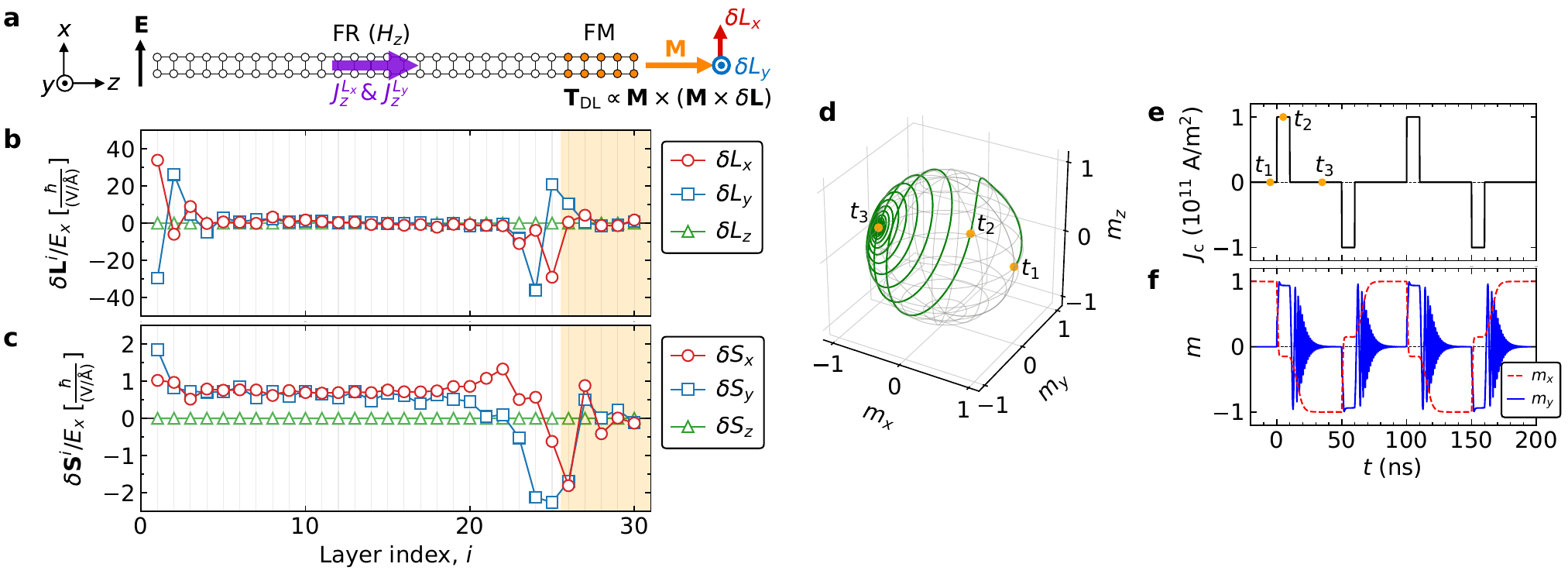}
	\caption{Unconventional orbital torque in a FR/FM bilayer and its application to field-free magnetization switching. \textbf{a} Crystal structure of the tight-binding model for a FR/FM bilayer. The FR bulk with $\hat{H}_z$ generates the conventional and unconventional orbital Hall currents, leading to orbital accumulation $\delta L_x$ and $\delta L_y$ at boundaries, respectively. The induced spin $\delta \mathbf{S} \propto \mathbf{M} \times \delta \mathbf{L}$ in the FM layer acts as an effective field that gives rise to the damping-like torque $\mathbf{T}_\mathrm{DL} \propto \mathbf{M} \times (\mathbf{M} \times \delta \mathbf{L})$. \textbf{b, c} Layer-resolved induced \textbf{b} OAM and \textbf{c} spin under an applied electric field $\mathbf{E} = E_x \hat{\mathbf{x}}$. \textbf{d} Macrospin simulation of type-$x$ magnetization switching with conventional and unconventional damping-like orbital torques, described by Eq.~\eqref{eq:llg}. During the time $t_1 \rightarrow t_2 \rightarrow t_3$ shown in \textbf{e}, the magnetization direction switches from $\hat{\mathbf{x}}$ to $-\hat{\mathbf{x}}$ upon a current pulse $J_\mathrm{c} = 10^{11} \, \mathrm{A/m}^2$ without external magnetic field. \textbf{e} Applied charge current density as a function of time $t$. \textbf{f} Dynamics of $m_x$ and $m_y$ in the same time domain.}
	\label{fig5} 
\end{figure*}

\subsection{Unconventional orbital torque and field-free switching}

So far, we have discussed rotation-induced orbital currents based on the conventional definition of the orbital current operator, which is not directly measurable. In this section, we show that FR order gives rise not only to orbital currents but also to effects that can be probed experimentally, such as OAM accumulation and orbital torque. To illustrate this, we examine a FR/FM bilayer using a tight-binding model (Fig.~\ref{fig5}a; see Methods). In our model, under $\mathbf{E} = E_x \hat{\mathbf{x}}$, the FR layer with $\hat{H}_z$ generates conventional ($J_z^{L_y}$) and unconventional ($J_z^{L_x}$) orbital Hall currents without spin currents, while the FM layer does not produce orbital or spin Hall currents on its own. This design ensures that the current-induced torque on magnetization $\mathbf{M} $ in the FM layer originates solely from the OAM injection by the FR layer. 

Within linear-response theory, we compute the current-induced non-equilibrium OAM ($\delta \mathbf{L}$) and spin ($\delta \mathbf{S}$) in the FR/FM system with $\hat{\mathbf{m}} = \hat{\mathbf{z}}$, where $\hat{\mathbf{m}}$ is the unit vector of $\mathbf{M}$ (Methods). Figure~\ref{fig5}b shows the layer-resolved $\delta \mathbf{L}$ per applied electric field. Large OAM components along $x$ and $y$ appear near the top and bottom surfaces of the FR layer, demonstrating orbital accumulation from the unconventional ($\delta L_x$) and conventional ($\delta L_y$) orbital Hall currents. The induced OAM is transferred across the FR/FM interface and subsequently interacts with $\mathbf{M}$. This generates $\delta \mathbf{S}$, which acts as an effective field for the torque $\mathbf{T} \propto \hat{\mathbf{m}} \times \delta \mathbf{S}$~\cite{go2020orbital, Salemi2021}. In particular, the spin-orbit precession with $\delta \mathbf{L}$ results in $\delta \mathbf{S} \propto \hat{\mathbf{m}} \times \delta \mathbf{L}$~\cite{go2023long}, leading to the damping-like orbital torque $\mathbf{T}_\mathrm{DL} \propto \hat{\mathbf{m}} \times (\hat{\mathbf{m}} \times \delta \mathbf{L})$. Due to the presence of  $\delta L_y$ and $\delta L_x$, we obtain $\mathbf{T}_\mathrm{DL} = \hat{\mathbf{m}} \times (\mathbf{B}_\mathrm{DL}^\mathrm{OT} + \mathbf{B}_\mathrm{DL}^\mathrm{UOT}) $, where $\mathbf{B}_\mathrm{DL}^\mathrm{OT} =  J_\mathrm{c} b_\mathrm{DL}^\mathrm{OT} \hat{\mathbf{m}} \times \hat{\mathbf{y}}$ and $\mathbf{B}_\mathrm{DL}^\mathrm{UOT}  = J_\mathrm{c} b_\mathrm{DL}^\mathrm{UOT} \hat{\mathbf{m}} \times \hat{\mathbf{x}}$ are the effective fields for the conventional and unconventional orbital torques, respectively. The effective fields per charge current density $J_\mathrm{c}$ are estimated in our model (see Methods) as $b_\mathrm{DL}^\mathrm{OT}  =  14 \ \mathrm{mT}/(10^{11} \, \mathrm{A}/\mathrm{m}^2) $ and $b_\mathrm{DL}^\mathrm{UOT}  =  -2 \ \mathrm{mT}/(10^{11} \, \mathrm{A}/\mathrm{m}^2) $, which fall within the range of reported values for spin-orbit torque devices using heavy metals~\cite{manchon2019current}. 

To gain further insight into how these orbital torques contribute to magnetization switching, we simulate magnetization dynamics within a macrospin model. The dynamics of $\hat{\mathbf{m}} = (m_x, m_y, m_z)$ is described by the Landau-Lifshitz-Gilbert equation
\begin{equation}\label{eq:llg}
	\frac{d\hat{\mathbf{m}}}{dt} = -\gamma \hat{\mathbf{m}} \times (\mathbf{B}_\mathrm{ani} + \mathbf{B}_\mathrm{DL}^\mathrm{OT} + \mathbf{B}_\mathrm{DL}^\mathrm{UOT}) + \alpha \hat{\mathbf{m}} \times \frac{d\hat{\mathbf{m}}}{dt}, 
\end{equation}
where $\gamma$ is the gyromagnetic ratio, $\mathbf{B}_\mathrm{ani}$ is the magnetic anisotropy field, and $\alpha$ is the Gilbert damping parameter. We consider a type-$x$ geometry~\cite{fukami2016spin}, in which an easy axis is collinear with the charge current, i.e., $\mathbf{B}_\mathrm{ani} = B_\mathrm{ani} m_x \hat{\mathbf{x}}$. Parameters are set to $\alpha = 0.05$ and $B_\mathrm{ani} = 30 \ \mathrm{mT}$. The same $\mathbf{B}_\mathrm{DL}^\mathrm{OT}$ and $\mathbf{B}_\mathrm{DL}^\mathrm{UOT}$ obtained above are used here, together with current pulses of $J_\mathrm{c} = \pm 10^{11} \, \mathrm{A}/\mathrm{m}^2$. Figure~\ref{fig5}d shows the trajectory of $\hat{\mathbf{m}}$ over time $t$ upon a current pulse, illustrated in Fig.~\ref{fig5}e. Initially, $\hat{\mathbf{m}}$ points along $\hat{\mathbf{x}}$ ($t=t_1$). The current pulse exerts the damping-like orbital torques on the magnetization. Because of the presence of $\mathbf{B}_\mathrm{DL}^\mathrm{UOT}$, $\hat{\mathbf{m}}$ is not perfectly aligned with $\hat{\mathbf{y}}$ but instead slightly tilted toward $-\hat{\mathbf{x}}$ ($t=t_2$), as shown in Fig.~\ref{fig5}f. Hence, after the pulse is turned off, $\hat{\mathbf{m}}$ relaxes deterministically to $- \hat{\mathbf{x}}$ ($t=t_3$) without the aid of an external magnetic field. Figures~\ref{fig5}e  and \ref{fig5}f illustrate the repetitive switching between $\pm \hat{\mathbf{x}}$ under opposite current pulses. These results demonstrate that unconventional orbital torques from FR materials offer a viable route to field-free magnetization switching.

\section{Discussion}

Unconventional orbital currents can emerge in general FR materials. To see this, we performed a symmetry analysis and determined the allowed components of the orbital (or spin) conductivity tensor $\bm{\sigma}^\mathbf{X}$ for all 32 crystallographic point groups (see Supplementary Note 4 and Table S1). We find that a total of 16 point groups allow unconventional components. These include all seven FR point groups that permit an axial vector order parameter, i.e., an electric toroidal dipole moment: $\bar{1}$, $2/m$, $\bar{3}$, $\bar{4}$, $\bar{6}$, $4/m$, and $6/m$~\cite{johnson2011cu}. In addition, their subgroups, 1, 2, 3, 4, 6, and $m$, also exhibit FR order and unconventional components. The remaining three point groups---$\bar{3}m$ and its subgroups $3m$ and $32$---do not allow FR order. This observation implies that unconventional orbital currents can also arise from other mechanisms not associated with the FR order, motivating further investigation into the microscopic origin in these systems.

It is also instructive to compare FR order with chiral order, which may appear similar. Both involve atomic rotations and lack vertical mirror planes, while chiral order additionally breaks inversion symmetry. Among the 11 chiral point groups, only six---1, 2, 4, 3, 6, and 32---support unconventional orbital currents, of which the first five are subgroups of FR point groups. This indicates that FR order plays a more fundamental role than chirality in generating unconventional orbital currents.

Finally, we note a possible connection between rotation-induced orbital currents and chirality-induced spin selectivity (CISS), a phenomenon where electrons acquire a net spin polarization upon transmission through a chiral material, acting as a spin filter~\cite{ray1999asymmetric, bloom2024chiral}. Longitudinal orbital currents in FR systems, for example $J_z^{L_z} = \sigma_{zz}^{L_z} E_z$, imply that such systems can act as orbital filters (and spin filters, when SOC is considered). Although a net OAM polarization is forbidden in bulk FR materials due to inversion symmetry, it can locally emerge at interfaces with local inversion symmetry breaking, as shown in Figs.~\ref{fig5}b and \ref{fig5}c. As a result, when asymmetric contacts are introduced, FR materials can induce a net OAM polarization, e.g., $L_z = \chi_{z}^{L_z} E_z$. This is consistent with the fact that removing inversion symmetry from FR point groups yield chiral point groups. Therefore, effects analogous to the CISS, or chirality-induced orbital selectivity~\cite{gobel2025chirality}, may also emerge in FR materials, opening a promising direction beyond conventional chiral systems.

To summarize, we have demonstrated that rotation-induced orbital currents can be electrically generated in FR systems and utilized for field-free orbital torque switching. The underlying mechanism originates from intrinsic, nonrelativistic orbital dynamics driven by an electric hexadecapole moment. Our symmetry analysis further suggests that a broad class of FR materials can exhibit these effects. Considering their phenomenological similarity to magnetization-induced ($\mathcal{T}$-odd) spin currents in FM systems, analogous functionalities---such as spin-transfer torque~\cite{ralph2008spin}---may be realized using rotation-induced ($\mathcal{T}$-even) orbital currents in FR systems, without requiring heavy elements. Consequently, FR order or higher-order electric multipoles may provide a fertile platform for exploring novel orbital transport phenomena, thereby expanding the landscape of orbitronics research.

\section{Methods}

\subsection{Three-dimensional tight-binding model}
We constructed a three-dimensional FR model (Fig.~\ref{fig3}a) with a tetragonal unit cell consisting of two atomic species: A with five $d$ orbitals ($d_{xy}, d_{yz}, d_{zx}, d_{x^2-y^2}, d_{3z^2 - r^2}$), and B with an $s$ orbital. The rotational displacement of the four B atoms, located at $\pm \frac{a}{3}( \cos \phi, \sin \phi,0)$ and $ \pm \frac{a}{3}( \sin \phi, - \cos \phi,0)$, induces the FR order along the $z$-axis. The lattice constants were set to $a=c=5$~\r{A}. SOC is neglected in this model. 

The tight-binding Hamiltonian of this system reads:
\begin{align}\label{eq:H_3d}
	\hat{\mathcal{H}}  = & \sum_{\langle i,j \rangle , m n \sigma } 	t_{ij,mn}^d \hat{d}_{im\sigma}^\dagger \hat{d}_{jn\sigma} +	\sum_{\langle \langle i,j \rangle \rangle , mn \sigma } 	t_{ij,mn}^{d(2)} \hat{d}_{im\sigma}^\dagger \hat{d}_{jn\sigma} \nonumber \\
	& +	\sum_{\langle l,j \rangle , n\sigma } 	t_{lj,n}^{sd} \hat{s}_{l\sigma}^\dagger \hat{d}_{jn\sigma} 
	+	\epsilon_{s}  \sum_{l\sigma} \hat{s}_{l\sigma}^\dagger \hat{s}_{l\sigma} ,
\end{align}
where $\hat{d}_{im\sigma}^\dagger$ and $\hat{d}_{im\sigma}$ ($\hat{s}_{l\sigma}^\dagger$ and $\hat{s}_{l\sigma}$) are the creation and annihilation operators for a $d$-electron ($s$-electron) at site $i$ on atom $A$ ($l$ on atom $B$), $\sigma = \uparrow, \downarrow$ denotes the spin, and $m$ specifies $d$-orbital species. Our model includes three types of electron hopping, as depicted in Fig.~\ref{fig3}a. The first two involve $d$-electron hopping between $A$ atoms: nearest-neighbor and next-nearest-neighbor hopping, with amplitudes $t_{ij,mn}^d$ and $t_{ij,mn}^{d(2)}$, respectively. The Slater-Koster parameters for the former are chosen as $(t_\sigma^d, t_\pi^d, t_\delta^d) = (-0.5, 0.2, -0.1)$ in eV, while those for the latter are scaled by a factor $\eta$, i.e., $(t_\sigma^{d(2)}, t_\pi^{d(2)}, t_\delta^{d(2)}) = (\eta t_\sigma^d, \eta t_\pi^d, \eta t_\delta^d)$. The third type of hopping occurs between nearest-neighbor $A$ and $B$ atoms via $sd$ hybridization with strength $t_\sigma^{sd} = -1.0$~eV. The corresponding hopping amplitude $t_{lj,n}^{sd}$ depends on $\phi$, thereby incorporating the FR order. The on-site energy for the $s$ electron is set to $\epsilon_{s} = -3$~eV. The momentum-space representation of the Hamiltonian is provided in Supplementary Note 1.

\subsection{Kubo formula for orbital conductivity}

The orbital conductivity was calculated by the Kubo formula~\cite{go2018intrinsic}:
\begin{align}
	\sigma_{\alpha \beta}^{L_\gamma} = &
	-e\hbar \int \frac{d^3\mathbf{k}}{(2\pi)^3} \sum_{n \neq m} 
	(f_{n\mathbf{k}} - f_{m\mathbf{k}}) \nonumber \\ 
	& \times  \mathrm{Im}
	\left[
	\frac
	{\langle \psi_{n\mathbf{k}} \vert \hat{J}_\alpha^{L_\gamma}  \vert \psi_ {m\mathbf{k}}\rangle 
		\langle \psi_{m\mathbf{k}} \vert \hat{v}_\beta \vert \psi_{n\mathbf{k}} \rangle 
	}
	{(\epsilon_{n\mathbf{k}} - \epsilon_{m\mathbf{k}})( \epsilon_{n\mathbf{k}} - \epsilon_{m\mathbf{k}} + i\Gamma )} 
	\right], \label{eq:kubo}
\end{align}
where $\vert \psi_{n\mathbf{k}} \rangle$ is the eigenstate, $\epsilon_{n\mathbf{k}}$ is the energy eigenvalue, $f_{n\mathbf{k}} = 1/[e^{ (\epsilon_{n\mathbf{k}} - \mu ) / k_\mathrm{B}T} + 1 ]$ is the Fermi-Dirac distribution function, with the chemical potential $\mu$ and $k_\mathrm{B}T = 25$~meV, $\hat{J}_\alpha^{L_\gamma} = \frac{1}{2} (\hat{v}_\alpha \hat{L}_\gamma +  \hat{L}_\gamma \hat{v}_\alpha) $ is the conventional orbital current operator, and $\hat{v}_\alpha $ is the $\alpha$-component of the velocity operator, evaluated by $\hat{v}_\alpha = \frac{1}{i\hbar} [\hat{r}_\alpha, \hat{\mathcal{H}}]$ with the position operator $\hat{r}_\alpha$. The lifetime broadening with $\Gamma =0.1$~eV accounts for scattering effects.

\subsection{First-principles calculation for TiAu$_4$}
 
 The electronic structure of the FR material TiAu$_4$ was calculated using the density functional theory code \texttt{FLEUR}~\cite{fleurWeb, fleurCode} based on the full-potential linearized augmented plane-wave method~\cite{wimmer1981full}. The Perdew-Burke-Ernzerhof functional~\cite{perdew1996generalized} within the generalized gradient approximation was used for an exchange-correlation functional. The conventional unit cell of tetragonal TiAu$_4$ (space group $I4/m$), with lattice parameters $a=6.55$~\r{A} and $c=3.96$~\r{A}, contains two Ti atoms and eight Au atoms located at the Wyckoff positions $2a$ and $8h$, respectively~\cite{jain2013commentary}. The muffin-tin radii were set to 2.71~$a_0$ and 2.57~$a_0$ for Ti and Au atoms, respectively, and the plane-wave cutoff was set to 3.8~$a_0^{-1}$, where $a_0$ is the Bohr radius. The SOC was included within the second variation scheme~\cite{li1990magnetic}. The Brillouin zone for the self-consistent calculation was sampled by a $9 \times 9 \times 9$ Monkhorst-Pack $\mathbf{k}$-mesh~\cite{monkhorst1976special}. 
 
 After the self-consistent calculation converged, we obtained the maximally localized Wannier functions (MLWFs)~\cite{freimuth2008maximally} using the \texttt{WANNIER90} code~\cite{pizzi2020wannier90}. Starting from the initial functions of $s$, $p_x$, $p_y$, $p_z$, $d_{xy}$, $d_{yz}$, $d_{zx}$, $d_{x^2-y^2}$, and $d_{3z^2-r^2}$ atomic orbitals for each spin (18 orbitals per atom), we constructed a total of 180 MLWFs, denoted as $\ket{n\mathbf{R}}$, where $\mathbf{R}$ is the lattice vector and $n$ is the band index. The frozen window energy maximum for the disentanglement was set to 12~eV above the Fermi energy, and a uniform $8 \times 8 \times 8$ $\mathbf{k}$-mesh was used. From the obtained MLWFs, the Hamiltonian ($\hat{\mathcal{H}}$), spin ($\hat{\mathbf{S}}$), OAM ($\hat{\mathbf{L}}$), and position ($\hat{\mathbf{r}}$) operators were given in the real-space representation, where the OAM was treated within the atom-centered approximation. Specifically, we obtained the matrix elements of the operators  $\hat{\mathcal{O}} = \hat{\mathcal{H}}$, $\hat{\mathbf{S}}$, $\hat{\mathbf{L}}$, and $\hat{\mathbf{r}}$ between the MLWFs, i.e., $\bra{m \mathbf{0}} \hat{\mathcal{O}} \ket{n \mathbf{R}}$. Defining the Bloch-like states in the Wannier gauge as
 \begin{equation}
 	\ket{\psi_{n\mathbf{k}}^\mathrm{W}} = \frac{1}{\sqrt{N}} \sum_\mathbf{R} e^{i \mathbf{k} \cdot \mathbf{R}} \ket{n \mathbf{R}} ,
 \end{equation}
 where $N$ is the number of the lattice vectors, we can construct the operators $\hat{\mathcal{O}}^\mathrm{W}(\mathbf{k})$ in the Wannier gauge on a dense $50 \times 50 \times 50$ $\mathbf{k}$-mesh, with the matrix elements given by:
 \begin{equation}
 	\mathcal{O}_{mn}^\mathrm{W}(\mathbf{k}) = \bra{\psi_{m\mathbf{k}}^\mathrm{W}} \hat{\mathcal{O}} \ket{\psi_{n\mathbf{k}}^\mathrm{W}} = \sum_\mathbf{R} e^{i \mathbf{k} \cdot \mathbf{R}} \bra{m \mathbf{0}} \hat{\mathcal{O}} \ket{n \mathbf{R}} .
 \end{equation}
 By diagonalizing the Hamiltonian $\hat{\mathcal{H}}^\mathrm{W}(\mathbf{k})$, we obtained the eigenstate $\ket{\psi_{n\mathbf{k}}}$ and the energy eigenvalue $\epsilon_{n\mathbf{k}}$ in the Hamiltonian gauge, satisfying $\hat{\mathcal{H}} \ket{\psi_{n\mathbf{k}}} = \epsilon_{n\mathbf{k}} \ket{\psi_{n\mathbf{k}}}$. Consequently, we computed the matrix elements for other operators, such as $\bra{\psi_{m\mathbf{k}}} \mathbf{S} \ket{\psi_{n\mathbf{k}}}$, $\bra{\psi_{m\mathbf{k}}} \mathbf{L} \ket{\psi_{n\mathbf{k}}}$, and $\bra{\psi_{m\mathbf{k}}} \mathbf{r} \ket{\psi_{n\mathbf{k}}}$, which were then used to evaluate the Kubo formula [Eq.~\eqref{eq:kubo}].

\subsection{Tight-binding calculation for a FR/FM bilayer}

The FR/FM bilayer model consists of square-lattice planes---25 FR and 5 FM monolayers---stacked along the $z$ axis. The system is periodic in the $xy$-plane and each layer contains a single atom with five $d$ orbitals ($d_{xy}, d_{yz}, d_{zx}, d_{x^2-y^2}, d_{3z^2 - r^2}$) per spin. The tight-binding Hamiltonian is written as
\begin{align}\label{eq:H_bilayer}
	\hat{\mathcal{H}} 
	 = &  \sum_{\langle i,j \rangle , mn\sigma } t_{ij,mn}^d \hat{d}_{im\sigma}^\dagger \hat{d}_{jn\sigma} 
	+ \sum_{\langle \langle i,j \rangle \rangle, mn \sigma }  t_{ij,mn}^{d(2)} \hat{d}_{im\sigma}^\dagger \hat{d}_{jn\sigma} \nonumber \\	
	& + \hat{\mathcal{H}}_\mathrm{FR} +  \hat{\mathcal{H}}_\mathrm{FM} .
\end{align}
The first two terms describe electron hopping, while $\hat{\mathcal{H}}_\mathrm{FR}$ and $\hat{\mathcal{H}}_\mathrm{FM}$ act only on the FR and FM layers, respectively, and are given by
\begin{subequations}
	\begin{align}
	\hat{\mathcal{H}}_\mathrm{FR} 
	 = &  \Delta_\mathrm{FR}  \sum_{i \in \mathrm{FR}}\sum_{mn \sigma} \hat{d}_{i m\sigma}^\dagger (\hat{H}_z)_{mn} \hat{d}_{i n\sigma}, \label{eq:H_FR} \\
	\hat{\mathcal{H}}_\mathrm{FM} 
	 = & \sum_{i \in \mathrm{FM}} (  \frac{\Delta_\mathrm{xc}}{\hbar} \sum_{n \sigma \sigma' }  \hat{d}_{i n\sigma}^\dagger \hat{ \mathbf{m} } \cdot\hat{\mathbf{S}}_{\sigma \sigma'} \; \hat{d}_{i n\sigma'}   \nonumber \\
	& +  \frac{\lambda_\mathrm{SO}}{\hbar^2} \sum_{ mn\sigma \sigma' } 
	\hat{d}_{i m\sigma}^\dagger \hat{\mathbf{L}}_{mn} \cdot \hat{\mathbf{S}}_{\sigma \sigma'} \; \hat{d}_{in\sigma'} )  . \label{eq:H_FM}
\end{align}
\end{subequations}
The FR order is implemented by the electric hexadecapole term with $\Delta_\mathrm{FR} = 0.2$~eV instead of explicitly modeling atomic rotations. Because $	\hat{\mathcal{H}}_\mathrm{FR} $ does not include SOC, the FR bulk generates orbital but not spin currents. The FM layer contains a local exchange field with $\Delta_\mathrm{xc} = 0.5$~eV and $\hat{\mathbf{m}}=\hat{\mathbf{z}}$, together with SOC of strength $\lambda_\mathrm{SO} = 0.05$~eV. The hopping amplitudes in Eq.~\eqref{eq:H_bilayer} are determined by the same Slater-Koster parameters used in the three-dimensional model [Eq.~\eqref{eq:H_3d}]: $(t_\sigma^d, t_\pi^d, t_\delta^d) = (-0.5, 0.2, -0.1)$ eV and $(t_\sigma^{d(2)}, t_\pi^{d(2)}, t_\delta^{d(2)}) = (\eta t_\sigma^d, \eta t_\pi^d, \eta t_\delta^d)$. We set $\eta=0.5$ for FR layers, $\eta=0$ for FM layers, and $\eta=0.25$ for interlayer hopping between FR and FM layers. Thus, the FM bulk lacks orbital texture and does not generate orbital or spin Hall currents.

The current-induced OAM ($\delta \mathbf{L}$) and spin ($\delta \mathbf{S}$) densities were calculated within linear-response formalism. Because $\mathcal{T}$ symmetry is broken in this system, both $\mathcal{T}$-even and $\mathcal{T}$-odd contributions to $\delta \mathbf{X}$ are allowed ($X = L$ or $S$), such that $\delta \mathbf{X} = \delta \mathbf{X}_\mathrm{even} + \delta \mathbf{X}_\mathrm{odd}$. In the small-$\Gamma$ limit, the $\mathcal{T}$-even and $\mathcal{T}$-odd contributions at the $i$-th atomic layer are given by~\cite{freimuth2014spin}:
\begin{subequations}
\begin{align}
	\frac{\delta \mathbf{X}^i_\mathrm{even}}{E_x} = &
	\frac{e\hbar}{\Gamma} \int \frac{d^2\mathbf{k}}{(2\pi)^2} \sum_{n} 
	(\frac{\partial f_{n\mathbf{k}}}{\partial \epsilon_{n\mathbf{k}}} ) \nonumber \\ 
	& \times \langle \psi_{n\mathbf{k}} \vert \hat{\mathbf{X}}^i  \vert \psi_ {n\mathbf{k}}\rangle 
		\langle \psi_{n\mathbf{k}} \vert \hat{v}_x \vert \psi_{n\mathbf{k}} \rangle , \label{eq:orb_den} \\
	\frac{\delta \mathbf{X}^i_\mathrm{odd}}{E_x} = &
	-e\hbar \int \frac{d^2\mathbf{k}}{(2\pi)^2} \sum_{n \neq m} 
	(f_{n\mathbf{k}} - f_{m\mathbf{k}}) \nonumber \\ 
	& \times  \mathrm{Im}
	\left[
	\frac
	{\langle \psi_{n\mathbf{k}} \vert \hat{\mathbf{X}}^i  \vert \psi_ {m\mathbf{k}}\rangle 
		\langle \psi_{m\mathbf{k}} \vert \hat{v}_x \vert \psi_{n\mathbf{k}} \rangle 
	}
	{(\epsilon_{n\mathbf{k}} - \epsilon_{m\mathbf{k}})( \epsilon_{n\mathbf{k}} - \epsilon_{m\mathbf{k}} + i\Gamma )} 
	\right], \label{eq:spin_den}
\end{align}
\end{subequations}
where $\hat{\mathbf{X}}^i$ is the OAM or spin operator projected onto the $i$-th atomic layer. We set $\Gamma = 10$~meV and the chemical potential is taken as $\mu = 0.2$~eV. Note that the $\mathcal{T}$-even and $\mathcal{T}$-odd angular momentum responses are governed by intraband and interband contributions, respectively, in contrast to the angular momentum current response [e.g., Eq.~\eqref{eq:kubo}]. In our system, $\delta \mathbf{L}^i$ is dominated by $\delta \mathbf{L}^i_\mathrm{even}$, which reflects the $\mathcal{T}$-even orbital currents in the FR layer.

The net current-induced torque acting on the FM layer can be written as $\mathbf{T} = \frac{d\mathbf{S}}{dt} =  \frac{ \Delta_\mathrm{xc}}{\hbar} \hat{\mathbf{m}} \times \delta \mathbf{S}^\mathrm{FM}$~\cite{go2020orbital}, where $\delta \mathbf{S}^\mathrm{FM} = \sum_{i \in \mathrm{FM}} \delta \mathbf{S}^i$ denotes the induced spin summed over the FM region. Accordingly, the effective field for $\mathbf{T} = \mathbf{m} \times \mathbf{B}$ is obtained by $\mathbf{B}  = \frac{\Delta_\mathrm{xc}}{\hbar m_\mathrm{s}} \delta \mathbf{S}^\mathrm{FM}$, where $\mathbf{m} = m_\mathrm{s} \hat{\mathbf{m}}$ with $m_\mathrm{s} = 7.0 \, \mu_\mathrm{B}$ is the total magnetic moment of the FM atoms in the unit cell. Notably, the damping-like orbital torque $\mathbf{T}_\mathrm{DL} \propto \hat{\mathbf{m}} \times (\hat{\mathbf{m}} \times \delta \mathbf{L})$ is associated with the $\mathcal{T}$-odd part of the induced spin, $\delta \mathbf{S}_\mathrm{odd} \propto \hat{\mathbf{m}} \times \delta \mathbf{L}_\mathrm{even}$. Hence, we estimated the effective field for the damping-like orbital torque by taking the $\mathcal{T}$-odd term, i.e., $\mathbf{B}_\mathrm{DL}  = \frac{\Delta_\mathrm{xc}}{\hbar m_\mathrm{s}} \delta \mathbf{S}_\mathrm{odd}^\mathrm{FM}$. The $x$ and $y$ components of $\delta \mathbf{S}_\mathrm{odd}^\mathrm{FM} $ per $E_x = 1 \, \textrm{V/\AA}$ are obtained as $-1.14 \, \hbar$ and  $-0.139 \, \hbar$, yielding $\mathbf{B}_\mathrm{DL}^\mathrm{OT} =  J_\mathrm{c} b_\mathrm{DL}^\mathrm{OT} \hat{\mathbf{m}} \times \hat{\mathbf{y}}$ and $\mathbf{B}_\mathrm{DL}^\mathrm{UOT}  = J_\mathrm{c} b_\mathrm{DL}^\mathrm{UOT} \hat{\mathbf{m}} \times \hat{\mathbf{x}}$, which correspond to the conventional and unconventional orbital torque contributions, respectively. The coefficients $b_\mathrm{DL}^\mathrm{OT}  =  14 \ \mathrm{mT}/(10^{11} \, \mathrm{A}/\mathrm{m}^2) $ and $b_\mathrm{DL}^\mathrm{UOT}  =  -2 \ \mathrm{mT}/(10^{11} \, \mathrm{A}/\mathrm{m}^2) $ are obtained by using a typical charge conductivity $\sigma_\mathrm{c} = 10^6 \ \Omega^{-1} \mathrm{m}^{-1}$ for $J_\mathrm{c} = \sigma_\mathrm{c} E_x$.

\section*{{Acknowledgements}}
We acknowledge Sang-Wook Cheong and Hyun-Woo Lee for insightful discussions. This work was supported by the Swedish Research Council (VR), the Knut and Alice Wallenberg Foundation (Grants No.\ 2022.0079 and No.\ 2023.0336), and the Wallenberg Initiative Materials Science for Sustainability (WISE) funded by the Knut and Alice Wallenberg Foundation. We further acknowledge support from the EIC Pathfinder OPEN grant No.\ 101129641 “OBELIX”. The calculations were supported by resources provided by the National Academic Infrastructure for Supercomputing in Sweden (NAISS) at NSC Link\"oping, partially funded by VR through Grant No.\ 2022-06725. 

\section*{{Author contributions}}
D.J. conceived the idea and performed the theoretical calculations under the supervision of P.M.O. Both authors wrote the manuscript.

\section*{{Competing interests}}
The authors declare no competing interests.

 \bibliography{ref.bib}

\end{document}


\title{Supplementary Information for "Unconventional orbital currents and torques due to ferro-rotational orbital textures"}
	

	\author{Daegeun~Jo}
	\email{daegeun.jo@physics.uu.se}
	\affiliation{Department of Physics and Astronomy, Uppsala University, P.O. Box 516, SE-75120 Uppsala, Sweden}
	\affiliation{Wallenberg Initiative Materials Science for Sustainability, Uppsala University, SE-75120 Uppsala, Sweden}
	
	\author{Peter~M.~Oppeneer}
	\email{peter.oppeneer@physics.uu.se}
	\affiliation{Department of Physics and Astronomy, Uppsala University, P.O. Box 516, SE-75120 Uppsala, Sweden}
	\affiliation{Wallenberg Initiative Materials Science for Sustainability, Uppsala University, SE-75120 Uppsala, Sweden}
	
	\maketitle
	
	\tableofcontents

	\section{Three-dimensional tight-binding model}\label{section1}
	
	The Hamiltonian in Eq.~(7) of the main text, which describes the three-dimensional tight-binding model, was formulated in the real-space representation. Here, we present its momentum-space representation in matrix form, which was used to obtain the Bloch eigenstates. First, we consider nine atomic orbital states denoted by $\ket{\phi_{n \mathbf{R}+\mathbf{r}_n}}$, where the orbital index $n$ corresponds to five $d$ orbitals---$d_{xy}$, $d_{yz}$, $d_{zx}$, $d_{x^2-y^2}$, and $d_{3z^2 - r^2}$---belonging to atom $A$, and four $s$ orbitals---denoted as $s_i$ ($i=1,2,3,4$)---for the $i$-th atom $B$. Here, $\mathbf{R}$ represents the lattice vector, and $\mathbf{r}_n $ represents the position of each orbital wave function center within the unit cell: $\mathbf{0} = (0,0,0)$ for $d$ orbitals and $\bm{\tau}_i$ for $s_i$ orbitals, where
	\begin{subequations}
		\begin{align}
			\bm{\tau}_1 & = \frac{a}{3}(\cos \phi, \sin \phi,0), \\
			\bm{\tau}_2 & = \frac{a}{3}(-\sin \phi, \cos \phi,0), \\
			\bm{\tau}_3 & = \frac{a}{3}(-\cos \phi, -\sin \phi,0), \\
			\bm{\tau}_4 & = \frac{a}{3}(\sin \phi, -\cos \phi,0),
		\end{align}
	\end{subequations}
	as illustrated in Fig. 3a of the main text. The Bloch-like states are then constructed as 
	\begin{equation}
		\ket{\varphi_{n\mathbf{k}}} = \frac{1}{\sqrt{N}} \sum_\mathbf{R} e^{i \mathbf{k} \cdot (\mathbf{R} + \mathbf{r}_n)} \ket{\phi_{n \mathbf{R}+\mathbf{r}_n}}  ,
	\end{equation}
	with the crystal momentum $\mathbf{k}$. Using these states as a basis, we construct a $9 \times 9$ spinless Hamiltonian matrix $\hat{\mathcal{H}}(\mathbf{k}) $, which can be partitioned into $d$- and $s$-orbital blocks as follows:
	\begin{equation}\label{eq:H_sd_block}
		\hat{\mathcal{H}}(\mathbf{k}) = 
		\begin{pmatrix}
			\hat{\mathcal{H}}^{d}(\mathbf{k}) & \hat{\mathcal{H}}^{ds}(\mathbf{k}) \\
			\hat{\mathcal{H}}^{sd}(\mathbf{k}) & \hat{\mathcal{H}}^{s}(\mathbf{k})
		\end{pmatrix},
	\end{equation}
	where $\hat{\mathcal{H}}^{d}(\mathbf{k})$, $\hat{\mathcal{H}}^{s}(\mathbf{k})$, and $\hat{\mathcal{H}}^{ds}(\mathbf{k}) = (\hat{\mathcal{H}}^{sd}(\mathbf{k}))^\dagger$ are $5 \times 5$, $4 \times 4$, and $5 \times 4$ matrices, respectively, with the superscript indicating the orbital type. Since spin-orbit coupling (SOC) is not included in our model, the total Hamiltonian including spin is simply given by   $ \hat{\mathcal{H}}(\mathbf{k})\otimes  \hat{I}_{2\times2} $, where $\hat{I}_{2\times2}$ is the $2\times2$ identity matrix in the spin basis. Within this basis, the orbital angular momentum operator $\hat{\mathbf{L}} = (\hat{L}_x, \hat{L}_y, \hat{L}_z) $ can be expressed as
	\begin{equation}
		\hat{\mathbf{L}} = 
		\begin{pmatrix}
			\hat{\mathbf{L}}^{d} & 0 \\
			0 & 0
		\end{pmatrix}
		\otimes  \hat{I}_{2\times2} ,
	\end{equation}
	where $\hat{\mathbf{L}}^d = (\hat{L}_x^d, \hat{L}_y^d, \hat{L}_z^d) $ corresponds to the atomic orbital angular momentum operator in the $d$-orbital basis, with each component defined as
	\begin{equation}
		\hat{L}_x^d = \hbar
		\begin{pmatrix}
			0 & 0 & -i & 0 & 0 \\
			0 & 0 & 0 & -i & -\sqrt{3}i \\
			i & 0 & 0 & 0 & 0 \\
			0 & i & 0 & 0 & 0 \\
			0 & \sqrt{3}i & 0 & 0 & 0
		\end{pmatrix}, \
		\hat{L}_y^d = \hbar
		\begin{pmatrix}
			0 & i & 0 & 0 & 0 \\
			-i & 0 & 0 & 0 & 0 \\
			0 & 0 & 0 & -i & \sqrt{3}i \\
			0 & 0 & i & 0 & 0 \\
			0 & 0 & -\sqrt{3}i & 0 & 0 
		\end{pmatrix}, \
		\hat{L}_z^d = \hbar
		\begin{pmatrix}
			0 & 0 & 0 & 2i & 0 \\
			0 & 0 & i & 0 & 0 \\
			0 & -i & 0 & 0 & 0 \\
			-2i & 0 & 0 & 0 & 0 \\
			0 & 0 &0 & 0 & 0 
		\end{pmatrix}.
	\end{equation}

	We now present each matrix element of $\hat{\mathcal{H}}(\mathbf{k}) $, written as
	\begin{equation}\label{eq:H_element}
		\mathcal{H}_{mn}(\mathbf{k}) \equiv \bra{\varphi_{m\mathbf{k}}} \hat{\mathcal{H}} \ket{\varphi_{n\mathbf{k}}} 
		= \sum_\mathbf{R} e^{i\mathbf{k} \cdot (\mathbf{R} + \mathbf{r}_n - \mathbf{r}_m)}
		\bra{\phi_{m,\mathbf{r}_m}} \hat{\mathcal{H}} \ket{\phi_{n, \mathbf{R}+\mathbf{r}_n}}.
	\end{equation}
	On the right-hand side, the energy integral $\bra{\phi_{m,\mathbf{r}_m}} \hat{\mathcal{H}} \ket{\phi_{n, \mathbf{R}+\mathbf{r}_n}}$ can be expressed in terms of Slater-Koster parameters, which describe $\sigma$-, $\pi$-, and $\delta$-type hopping between given orbital types. Within the two-center approximation, $\bra{\phi_{m,\mathbf{r}_m}} \hat{\mathcal{H}} \ket{\phi_{n, \mathbf{R}+\mathbf{r}_n}}$ is written as a linear combination of the corresponding Slater-Koster parameters that depend only on the hopping distance $| \mathbf{R} + \mathbf{r}_n - \mathbf{r}_m |$, with coefficients given by functions of the directional cosines of $ \mathbf{R} + \mathbf{r}_n - \mathbf{r}_m $ (e.g., see Table 1 of Ref.~\cite{slater1954simplified} for explicit formulae). In our model, we consider the following Slater-Koster parameters: $t_\sigma^d$, $t_\pi^d$, $t_\delta^d$ for nearest-neighbor hopping between $d$ orbitals, $t_\sigma^{d(2)}$, $t_\pi^{d(2)}$, $t_\delta^{d(2)}$ for next-nearest-neighbor hopping between $d$ orbitals, and $t_\sigma^{sd}$ for nearest-neighbor hopping between $s$ and $d$ orbitals. Note that hopping between $s$ orbitals is not considered in our model; only the on-site energy $\epsilon_s$ is included. Therefore, the $s$-orbital block in Eq.~\eqref{eq:H_sd_block} is given by 
	\begin{equation}\label{eq:H_s}
		\hat{\mathcal{H}}^s = \epsilon_s \hat{I}^s,
	\end{equation}
	where $\hat{I}^s$ is the $4 \times 4$ identity matrix in the $s$-orbital basis.
	
	For the $d$-orbital block $\hat{\mathcal{H}}^d(\mathbf{k})$, the matrix elements of Eq.~\eqref{eq:H_element} between $d$ orbitals on atom $A$, with $\mathbf{r}_m = \mathbf{r}_n = \mathbf{0}$, are given by~\cite{slater1954simplified}, specifically,
	\begingroup
	\allowdisplaybreaks
	\begin{eqnarray}\label{eq:H_d}
		\mathcal{H}_{xy,xy}(\mathbf{k})
		& =&   \epsilon_{xy} + 2 t_\pi^d (\cos \xi_x + \cos \xi_y) + 2 t_\delta^d \cos \xi_z  \nonumber \\
		& & +  2(t_\pi^{d(2)} + t_\delta^{d(2)}) (\cos \xi_x  \cos \xi_z + \cos \xi_y \cos \xi_z)
		+ (3 t_\sigma^{d(2)} + t_\delta^{d(2)}) \cos \xi_x  \cos \xi_y , \nonumber \\
		\mathcal{H}_{xy,yz}(\mathbf{k})
		& =& -2(t_\pi^{d(2)} - t_\delta^{d(2)}) \sin \xi_x  \sin \xi_z , \nonumber \\
		\mathcal{H}_{xy,zx}(\mathbf{k})
		& =& -2(t_\pi^{d(2)} - t_\delta^{d(2)}) \sin \xi_y  \sin \xi_z , \nonumber \\
		\mathcal{H}_{xy,x^2-y^2}(\mathbf{k})
		& =& 0 , \nonumber \\
		\mathcal{H}_{xy,3z^2-r^2}(\mathbf{k})
		& =& \sqrt{3} (t_\sigma^{d(2)} - t_\delta^{d(2)}) \sin \xi_x  \sin \xi_y , \nonumber \\
		\mathcal{H}_{yz,yz}(\mathbf{k})
		& =&   \epsilon_{yz} + 2 t_\pi^d (\cos \xi_y + \cos \xi_z) + 2 t_\delta^d \cos \xi_x  \nonumber \\
		& & +  2(t_\pi^{d(2)} + t_\delta^{d(2)}) (\cos \xi_x  \cos \xi_z + \cos \xi_x \cos \xi_y)
		+ (3 t_\sigma^{d(2)} + t_\delta^{d(2)}) \cos \xi_y  \cos \xi_z , \nonumber \\
		\mathcal{H}_{yz,zx}(\mathbf{k})
		& =& -2(t_\pi^{d(2)} - t_\delta^{d(2)}) \sin \xi_x  \sin \xi_y , \nonumber \\
		\mathcal{H}_{yz,x^2-y^2}(\mathbf{k})
		& =& \frac{3}{2} ( t_\sigma^{d(2)} - t_\delta^{d(2)})  \sin \xi_y  \sin \xi_z  , \nonumber \\
		\mathcal{H}_{yz,3z^2-r^2}(\mathbf{k})
		& =& -\frac{\sqrt{3}}{2} (t_\sigma^{d(2)} - t_\delta^{d(2)}) \sin \xi_y  \sin \xi_z , \nonumber \\
		\mathcal{H}_{zx,zx}(\mathbf{k})
		& =&   \epsilon_{zx} + 2 t_\pi^d (\cos \xi_x + \cos \xi_z) + 2 t_\delta^d \cos \xi_y  \nonumber \\
		& & +  2(t_\pi^{d(2)} + t_\delta^{d(2)}) (\cos \xi_x  \cos \xi_y + \cos \xi_y \cos \xi_z)
		+ (3 t_\sigma^{d(2)} + t_\delta^{d(2)}) \cos \xi_x  \cos \xi_z , \nonumber \\
		\mathcal{H}_{zx,x^2-y^2}(\mathbf{k})
		& =& -\frac{3}{2} ( t_\sigma^{d(2)} - t_\delta^{d(2)})  \sin \xi_x  \sin \xi_z  , \nonumber \\
		\mathcal{H}_{zx,3z^2-r^2}(\mathbf{k})
		& =& -\frac{\sqrt{3}}{2} (t_\sigma^{d(2)} - t_\delta^{d(2)}) \sin \xi_x  \sin \xi_z , \nonumber \\
		\mathcal{H}_{x^2-y^2,x^2-y^2}(\mathbf{k})
		& =& \epsilon_{x^2-y^2} +   \frac{1}{2}(3 t_\sigma^d +  t_\delta^d) (\cos \xi_x + \cos \xi_y) + 2 t_\delta^d \cos \xi_z \nonumber \\
		& & +  (\frac{3}{4} t_\sigma^{d(2)} + t_\pi^{d(2)} + \frac{9}{4} t_\delta^{d(2)}) (\cos \xi_x \cos \xi_z + \cos \xi_y \cos \xi_z) + 4 t_\pi^{d(2)} \cos \xi_x \cos \xi_y , \nonumber \\
		\mathcal{H}_{x^2-y^2,3z^2-r^2}(\mathbf{k})
		& =&  -\frac{\sqrt{3}}{2}(t_\sigma^d - t_\delta^d ) (\cos \xi_x - \cos \xi_y) \nonumber	\\
		& & + (\frac{\sqrt{3}}{4} t_\sigma^{d(2)} - \sqrt{3} t_\pi^{d(2)} + \frac{3\sqrt{3}}{4} t_\delta^{d(2)}) ( \cos \xi_x \cos \xi_z - \cos \xi_y \cos \xi_z) , \nonumber \\
		\mathcal{H}_{3z^2-r^2,3z^2-r^2}(\mathbf{k})
		& =& \epsilon_{3z^2-r^2} + \frac{1}{2}(t_\sigma^d + 3 t_\delta^d ) (\cos \xi_x + \cos \xi_y) + 2 t_\sigma^d \cos \xi_z \nonumber \\
		& & +  (\frac{1}{4} t_\sigma^{d(2)} + 3 t_\pi^{d(2)} + \frac{3}{4} t_\delta^{d(2)}) (\cos \xi_x \cos \xi_z 	+ \cos \xi_y \cos \xi_z)  \nonumber \\
		& &  + ( t_\sigma^{d(2)} + 3 t_\delta^{d(2)} ) \cos \xi_x \cos \xi_y ,
	\end{eqnarray}
	\endgroup
	where $\xi_x = k_x a$, $\xi_y = k_y a$, and $\xi_z = k_z a$. Note that the on-site energies $\epsilon_{xy}$, $\epsilon_{yz}$, $\epsilon_{zx}$, $\epsilon_{x^2-y^2}$, and $\epsilon_{3z^2-r^2}$ for $d$ orbitals are assumed to be zero in the main text for simplicity. 
	
	For the off-diagonal block $\hat{\mathcal{H}}^{ds}(\mathbf{k}) = (\hat{\mathcal{H}}^{sd}(\mathbf{k}))^\dagger$, the matrix elements between $d$ and $s_i$ orbitals (on atom $A$ and $i$-th atom $B$, respectively) are given by
	\begin{eqnarray}\label{eq:H_sd}
		\mathcal{H}_{xy,s_i}(\mathbf{k}) & = & (-1)^{i+1}  \frac{\sqrt{3} t_\sigma^{sd}}{2} e^{i \mathbf{k} \cdot \bm{\tau}_i}  \sin 2\phi , \nonumber \\
		\mathcal{H}_{yz,s_i}(\mathbf{k}) & = &  0, \nonumber \\
		\mathcal{H}_{zx,s_i}(\mathbf{k}) & = &  0, \nonumber \\
		\mathcal{H}_{x^2-y^2,s_i}(\mathbf{k}) & = &  (-1)^{i+1}  \frac{\sqrt{3} t_\sigma^{sd}}{2} e^{i \mathbf{k} \cdot \bm{\tau}_i} \cos 2\phi , \nonumber \\
		\mathcal{H}_{3z^2-r^2,s_i}(\mathbf{k}) & = & -  \frac{ t_\sigma^{sd}}{2} e^{i \mathbf{k} \cdot \bm{\tau}_i} .
	\end{eqnarray}
	It is important to note that $\hat{\mathcal{H}}^{ds}(\mathbf{k})$ and $\hat{\mathcal{H}}^{sd}(\mathbf{k})$ are dependent on $\phi$. The $sd$-hybridization through these terms reflects the ferro-rotational order, making the $d$-orbital states have the electric hexadecapole moment, as will be demonstrated in \ref{section2}. 
	
	Consequently, diagonalizing $ \hat{\mathcal{H}}(\mathbf{k})\otimes  \hat{I}_{2\times2} $ with matrix elements given in Eqs.~\eqref{eq:H_s}--\eqref{eq:H_sd} yields the Bloch eigenstate $\ket{\psi_{n\mathbf{k}}}$ with the energy eigenvalue $\epsilon_{n\mathbf{k}}$ used in the main text.

	\section{Electric hexadecapole moment induced by ferro-rotational order}\label{section2}
	
	While it is known that an electric hexadecapole moment is compatible with the ferro-rotational (FR) order from a symmetry perspective~\cite{hayami2022electric, hayami2023unconventional}, their direct relationship has not been clearly established. In this section, starting from the $sd$-orbital tight-binding model introduced in \ref{section1}, with structural rotation by an angle $\phi$, we derive an effective Hamiltonian $\hat{\mathcal{H}}^{d,\mathrm{eff}}(\mathbf{k})$ by downfolding the full Hamiltonian into the $d$-orbital space. We find that the electric hexadecapole moment operator is naturally incorporated into $\hat{\mathcal{H}}^{d,\mathrm{eff}}(\mathbf{k})$ as a consequence of $\phi$-dependent electron hopping. This result reveals a direct connection between the FR order and electric hexadecapole moment, thereby validating the minimal model Hamiltonian given in Eq.~(3) of the main text.

	The FR order in our model arises from the rotational displacement of atoms within the $xy$-plane. Therefore, the $xy$-plane structure captures the essential features of the FR order. To simplify the problem, we project the three-dimensional system onto a two-dimensional model by considering a single atomic layer in the $xy$-plane of the original system (see the right panel in Fig.~3a of the main text). The resulting two-dimensional Hamiltonian retains the form of Eq.~\eqref{eq:H_sd_block}, but all $k_z$-dependent  terms in Eq.~\eqref{eq:H_d} vanish. For the sake of simplicity, we restrict hopping to nearest neighbors and neglect next-nearest-neighbor hopping, i.e., $t_\sigma^{d(2)} = t_\pi^{d(2)} = t_\delta^{d(2)} = 0$. As a result, the $d$-orbital block $\hat{\mathcal{H}}^d(\mathbf{k}) $ of the spinless Hamiltonian simplifies to
	\begin{equation}\label{eq:H_d_nearest}
		\hat{\mathcal{H}}^d(\mathbf{k}) = 
		\begin{pmatrix}
			\mathcal{H}_{xy}(\mathbf{k}) & 0 & 0 & 0 & 0 \\
			0 & \mathcal{H}_{yz}(\mathbf{k}) & 0 & 0 & 0 \\
			0 & 0 & \mathcal{H}_{zx}(\mathbf{k}) & 0 & 0 \\
			0 & 0 & 0 & \mathcal{H}_{x^2-y^2}(\mathbf{k})  & \gamma(\mathbf{k}) \\
			0 & 0 & 0 & \gamma(\mathbf{k})  & 	\mathcal{H}_{3z^2-r^2}(\mathbf{k})
		\end{pmatrix},
	\end{equation}
	where the matrix elements are given by
	\begin{subequations}
		\begin{eqnarray}
			\mathcal{H}_{xy}(\mathbf{k}) & = &  
			\epsilon_{xy} + 2 t_\pi^d (\cos k_xa + \cos k_ya)  , \label{eq:H_2d_xy} \\
			\mathcal{H}_{yz}(\mathbf{k}) & = &  
			\epsilon_{yz} +  2 t_\delta^d \cos k_xa + 2 t_\pi^d \cos k_ya  , \\
			\mathcal{H}_{zx}(\mathbf{k}) & = &  
			\epsilon_{zx} +  2 t_\pi^d \cos k_xa +  2 t_\delta^d \cos k_ya , \\
			\mathcal{H}_{x^2-y^2}(\mathbf{k}) & =  & 
			\epsilon_{x^2-y^2} +  (\frac{3}{2} t_\sigma^d + \frac{1}{2} t_\delta^d) (\cos k_xa + \cos k_ya) 	, \label{eq:H_2d_x2y2} \\
			\mathcal{H}_{3z^2-r^2}(\mathbf{k}) & = & 
			\epsilon_{3z^2-r^2} +  (\frac{1}{2} t_\sigma^d + \frac{3}{2} t_\delta^d) (\cos k_xa + \cos k_ya) , \label{eq:H_2d_z2} \\
			\gamma(\mathbf{k}) & = & 
			-\frac{\sqrt{3}}{2}(t_\sigma^d - t_\delta^d ) (\cos k_xa - \cos k_ya)	. \label{eq:H_2d_gamma}
		\end{eqnarray}
	\end{subequations}
	The other blocks $\hat{\mathcal{H}}^s$ and $\hat{\mathcal{H}}^{ds}$ remain unchanged, as given in Eqs.~\eqref{eq:H_s} and \eqref{eq:H_sd}, respectively. Specifically, the matrix representation of $\hat{\mathcal{H}}^{ds}$ reads
	\begin{equation}\label{eq:H_sd_matrix}
		\hat{\mathcal{H}}^{ds}(\mathbf{k}) =  	(\hat{\mathcal{H}}^{sd}(\mathbf{k}) )^\dagger = 
		\frac{\sqrt{3}t_\sigma^{sd}}{2}
		\begin{pmatrix}
			e^{i \mathbf{k} \cdot \bm{\tau}_1} \sin 2\phi & 
			-e^{i \mathbf{k} \cdot \bm{\tau}_2} \sin 2\phi & 
			e^{-i \mathbf{k} \cdot \bm{\tau}_1} \sin 2\phi & 
			-e^{-i \mathbf{k} \cdot \bm{\tau}_2} \sin 2\phi  \\ 
			0 & 0 & 0 & 0 \\
			0 & 0 & 0 & 0 \\
			e^{i \mathbf{k} \cdot \bm{\tau}_1} \cos 2\phi & 
			-e^{i \mathbf{k} \cdot \bm{\tau}_2} \cos 2\phi & 
			e^{-i \mathbf{k} \cdot \bm{\tau}_1} \cos 2\phi & 
			-e^{-i \mathbf{k} \cdot \bm{\tau}_2} \cos 2\phi  \\ 
			-\frac{1}{\sqrt{3}} e^{i \mathbf{k} \cdot \bm{\tau}_1}  & 
			-\frac{1}{\sqrt{3}} e^{i \mathbf{k} \cdot \bm{\tau}_2}  & 
			-\frac{1}{\sqrt{3}} e^{-i \mathbf{k} \cdot \bm{\tau}_1} & 
			-\frac{1}{\sqrt{3}} e^{-i \mathbf{k} \cdot \bm{\tau}_2} 
		\end{pmatrix}.
	\end{equation}
	It is important to note that all FR information is encoded in Eq.~\eqref{eq:H_sd_matrix}. To investigate its influence on the $d$-orbital electrons of atom $A$, we downfold the full Hamiltonian 
	$\hat{\mathcal{H}}(\mathbf{k}) = 
	\begin{pmatrix}
		\hat{\mathcal{H}}^{d}(\mathbf{k}) & \hat{\mathcal{H}}^{sd}(\mathbf{k}) \\
		(	\hat{\mathcal{H}}^{sd}(\mathbf{k}) )^\dagger & \hat{\mathcal{H}}^{s}(\mathbf{k})
	\end{pmatrix}$
	into the subspace spanned by $d$ orbitals. This is achieved by using the Löwdin downfolding technique~\cite{lowdin1962studies}, as follows:
	\begin{align}
		\hat{\mathcal{H}}^{d,\mathrm{eff}} (\mathbf{k}) & = \hat{\mathcal{H}}^d(\mathbf{k}) + \hat{\mathcal{H}}^{ds}(\mathbf{k}) (\epsilon \hat{I}^s - \hat{\mathcal{H}}^{s} )^{-1} \hat{\mathcal{H}}^{sd}(\mathbf{k}) \nonumber \\
		& = 	\hat{\mathcal{H}}^d(\mathbf{k})  +
		\frac{(t_\sigma^{sd})^2}{\epsilon - \epsilon_s} 
		\begin{pmatrix}
			3 \sin^2 2\phi & 0 & 0 & 3 \sin 2\phi \cos 2\phi & 0 \\
			0 & 0 & 0 & 0 & 0 \\
			0 & 0 & 0 & 0 & 0 \\
			3 \sin 2\phi \cos 2\phi  & 0 & 0 & 3 \cos^2 2\phi  & 0 \\
			0 & 0 & 0 & 0  & 1
		\end{pmatrix}, \label{eq:lowdin}
	\end{align}
	where $\epsilon$ represents the eigenvalue of the full Hamiltonian. Equation~\eqref{eq:lowdin} implies that the original Hamiltonian $\hat{\mathcal{H}}^d$ in the $d$-subspace is dressed with additional terms originating from $sd$-hopping. Since $\hat{\mathcal{H}}^{d,\mathrm{eff}}(\mathbf{k})$ depends on $\epsilon$, solving the nonlinear eigenvalue problem requires iterative calculations. However, as we are primarily interested in the qualitative properties of the effective Hamiltonian rather than exact eigenvalues, we treat $\epsilon$ as a constant. Assuming a small rotation angle $\phi $, the effective Hamiltonian is then approximated as	
	\begin{equation}\label{eq:H_d_eff}
		\hat{\mathcal{H}}^{d,\mathrm{eff}}(\mathbf{k}) \approx
		\hat{\mathcal{H}}^d(\mathbf{k})  + \hat{V} + \Delta (\phi) \hat{H}_z^d ,
	\end{equation}
	up to linear order in $\phi$. The second term on the right-hand side of Eq.~\eqref{eq:H_d_eff}, 
	\begin{equation}
		\hat{V} = \frac{(t_\sigma^{sd})^2}{\epsilon - \epsilon_s}
		\begin{pmatrix}
			0 & 0 & 0 & 0 & 0 \\
			0 & 0 & 0 & 0 & 0 \\
			0 & 0& 0 & 0 & 0 \\
			0 & 0 & 0 & 3 & 0 \\
			0 & 0 &0 & 0 & 1
		\end{pmatrix} ,
	\end{equation}
	simply amounts to a correction to the on-site energies for $d$ orbitals. More importantly, the third term corresponds to the electric hexadecapole moment operator [e.g., Eq. (2) of the main text] in the $d$-orbital basis,
	\begin{equation}
		\hat{H}_z^d = \frac{1}{12\hbar^4} \{ \{\hat{L}_x^d, \hat{L}_y^d \} , (\hat{L}_x^d)^2 - (\hat{L}_y^d)^2  \} = 
		\begin{pmatrix}
			0 & 0 & 0 & 1 & 0 \\
			0 & 0 & 0 & 0 & 0 \\
			0 & 0& 0 & 0 & 0 \\
			1 & 0 & 0 & 0 & 0 \\
			0 & 0 &0 & 0 & 0 
		\end{pmatrix},
		\label{eq:hexadecapole}
	\end{equation}
	with the coefficient
	\begin{equation}
		\Delta (\phi) = \frac{6(t_\sigma^{sd})^2}{\epsilon - \epsilon_s} \phi 
	\end{equation}
	being proportional to $\phi$. This result explicitly demonstrates that the electric hexadecapole moment naturally emerges from the FR order. Accordingly, the electric hexadecapole moment serves as a relevant parameter for describing the nonrelativistic physics of $d$-orbital FR systems. Note that the inclusion of SOC can introduce additional multipole moment operators defined in the spinful basis, such as (atomic-site) electric toroidal moments, which are known to generate relativistic spin phenomena, as demonstrated in Refs.~\cite{hayami2022electric, hayami2023unconventional}. In our work, however, we focus only on the nonrelativistic effects and ignore the relativistic contributions.
	
	The effective Hamiltonian in Eq.~\eqref{eq:H_d_eff} can be further simplified to
	\begin{equation}\label{eq:H_d_eff_2}
		\hat{\mathcal{H}}^{d,\mathrm{eff}}(\mathbf{k}) = 
		\begin{pmatrix}
			\mathcal{H}_{xy}(\mathbf{k}) & \Delta & 0 \\
			\Delta & \mathcal{H}_{x^2-y^2}(\mathbf{k}) & \gamma(\mathbf{k}) \\
			0 & \gamma(\mathbf{k})  & \mathcal{H}_{3z^2-r^2}(\mathbf{k})
		\end{pmatrix}	,
	\end{equation}
	written in the \{$d_{xy}, d_{x^2-y^2}, d_{3z^2-r^2}$\} basis, where $\Delta$ is now regarded as a constant. The $d_{yz}$ and $d_{zx}$ orbitals are excluded, as they do not couple to other orbitals, as evident from Eq.~\eqref{eq:H_d_nearest}, and $\hat{V}$ in Eq.~\eqref{eq:H_d_eff}  is neglected because it can be absorbed into $\mathcal{H}_{x^2-y^2}(\mathbf{k})$ and $\mathcal{H}_{3z^2-r^2}(\mathbf{k})$ by modifying the on-site energies $\epsilon_{x^2-y^2}$ and $\epsilon_{3z^2-r^2}$, respectively. Furthermore, one can assume that the coupling  between the $d_{x^2-y^2}$ and $d_{3z^2-r^2}$ orbitals, described by $\gamma(\mathbf{k})$, is sufficiently weak relative to the energy separation between their dispersions. For instance, a strong crystal field in a square planar structure can result in a large orbital splitting, satisfying $ \vert \epsilon_{x^2-y^2} - \epsilon_{3z^2-r^2} \vert \gg \vert \gamma \vert \sim \vert t_\sigma^d - t_\delta^d \vert$. In this case, the top-left $2 \times 2$ block in Eq.~\eqref{eq:H_d_eff_2}, which corresponds to Eq.~(3) of the main text, can effectively describe the FR system in the two-orbital basis \{$d_{xy}, d_{x^2-y^2}$\}.

	Within the assumption of small $\gamma$, a two-orbital effective Hamiltonian in the $d_{xy}$-$d_{x^2-y^2}$ subspace can be more formally derived using the Schrieffer-Wolff transformation~\cite{schrieffer1966relation}. The Hamiltonian in Eq.~\eqref{eq:H_d_eff_2}, hereafter referred to as $\hat{\mathcal{H}}$, can be separated into two parts:
	\begin{equation}
		\hat{\mathcal{H}} = \hat{\mathcal{H}}_0 + \hat{\mathcal{H}}_1 ,
	\end{equation}
	with
	\begin{equation}
		\hat{\mathcal{H}}_0 = 
		\begin{pmatrix}
			\mathcal{H}_{xy} & \Delta & 0 \\
			\Delta & \mathcal{H}_{x^2-y^2} & 0 \\
			0 & 0 & \mathcal{H}_{3z^2-r^2}
		\end{pmatrix} ,
		\qquad
		\hat{\mathcal{H}}_1 =  
		\begin{pmatrix}
			0 & 0 & 0 \\
			0 & 0 & \gamma \\
			0 & \gamma  & 0
		\end{pmatrix},
	\end{equation}
	where the explicit $\mathbf{k}$-dependence is omitted for brevity. Meanwhile, the generator $\hat{S}$ such that $[\hat{S},\hat{\mathcal{H}}_0] = - \hat{\mathcal{H}}_1$ is defined as
	\begin{equation}
		\hat{S} = \frac{\gamma}{(\mathcal{H}_{3z^2-r^2}-\mathcal{H}_{xy})(\mathcal{H}_{3z^2-r^2}-\mathcal{H}_{x^2-y^2}) - \Delta^2}
		\begin{pmatrix}
			0 & 0 & -\Delta \\
			0 & 0 & -(\mathcal{H}_{3z^2-r^2}-\mathcal{H}_{xy}) \\
			\Delta & \mathcal{H}_{3z^2-r^2}-\mathcal{H}_{xy} & 0
		\end{pmatrix}.
	\end{equation}
	The unitary transformation $\hat{\mathcal{H}}' = e^{\hat{S}} \hat{\mathcal{H}} e^{-\hat{S}}$  then diagonalizes $\hat{\mathcal{H}}$ in the eigenbasis of $\hat{\mathcal{H}}_0$ to first order in $\hat{\mathcal{H}}_1$, as follows:
	\begin{align}
		\hat{\mathcal{H}}' & = \hat{\mathcal{H}} + [\hat{S},\hat{\mathcal{H}}] + \frac{1}{2}[\hat{S},[\hat{S},\hat{\mathcal{H}}]] + \cdots \nonumber \\
		&= \hat{\mathcal{H}}_0 + \hat{\mathcal{H}}_1 + [\hat{S},\hat{\mathcal{H}}_0] + [\hat{S},\hat{\mathcal{H}}_1] + \frac{1}{2}[\hat{S},[\hat{S},\hat{\mathcal{H}}_0]] + \cdots \nonumber \\
		&= \hat{\mathcal{H}}_0 + \frac{1}{2}[\hat{S},\hat{\mathcal{H}}_1] + O(\hat{\mathcal{H}}_1^3) \nonumber \\
		& \approx 
		\hat{\mathcal{H}}_0 
		+
		\frac{\gamma^2}{(\mathcal{H}_{3z^2-r^2}-\mathcal{H}_{xy})(\mathcal{H}_{3z^2-r^2}-\mathcal{H}_{x^2-y^2}) - \Delta^2}
		\begin{pmatrix}
			0 & -\frac{1}{2}\Delta & 0 \\
			-\frac{1}{2}\Delta & -(\mathcal{H}_{3z^2-r^2}-\mathcal{H}_{xy}) & 0 \\
			0 & 0 & \mathcal{H}_{3z^2-r^2}-\mathcal{H}_{xy}
		\end{pmatrix}.
	\end{align}
	Consequently, the $d_{3z^2-r^2}$ subspace can be effectively decoupled from the $d_{xy}$-$d_{x^2-y^2}$ subspace, leading to the following effective Hamiltonian, to linear order in $\Delta$:
	\begin{equation}\label{eq:H_d_eff_3}
		\hat{\mathcal{H}}'(\mathbf{k}) = 
		\begin{pmatrix}
			\mathcal{H}_{xy}(\mathbf{k}) & \Delta'(\mathbf{k}) \\
			\Delta'(\mathbf{k}) & \mathcal{H}_{x^2-y^2}'(\mathbf{k}) 
		\end{pmatrix}	,
	\end{equation}
	where
	\begin{subequations}
		\begin{align}
			\Delta'(\mathbf{k}) & = \Delta \left[1 - \frac{\gamma(\mathbf{k})^2}
			{2(\mathcal{H}_{3z^2-r^2}(\mathbf{k}) - \mathcal{H}_{xy}(\mathbf{k}))
				(\mathcal{H}_{3z^2-r^2}(\mathbf{k}) - \mathcal{H}_{x^2-y^2}(\mathbf{k}))}
			\right]  , \\
			\mathcal{H}_{x^2-y^2}'(\mathbf{k}) & = \mathcal{H}_{x^2-y^2}(\mathbf{k}) \left[ 1 - 
			\frac{\gamma(\mathbf{k})^2}
			{\mathcal{H}_{x^2-y^2}(\mathbf{k})
				(\mathcal{H}_{3z^2-r^2}(\mathbf{k}) - \mathcal{H}_{x^2-y^2}(\mathbf{k}))} \right],
		\end{align}
	\end{subequations}
	with $\mathcal{H}_{xy}(\mathbf{k})$, $\mathcal{H}_{x^2-y^2}(\mathbf{k})$, $\mathcal{H}_{3z^2-r^2}(\mathbf{k})$, and $\gamma(\mathbf{k})$ given in Eqs.~\eqref{eq:H_2d_xy} \eqref{eq:H_2d_x2y2}, \eqref{eq:H_2d_z2}, and \eqref{eq:H_2d_gamma}, respectively. These results show that if the coupling strength between $d_{x^2-y^2}$ and $d_{3z^2-r^2}$ orbitals, $ \vert \gamma \vert \sim \vert t_\sigma^d - t_\delta^d \vert $, is sufficiently smaller than the crystal field splitting $ \vert \epsilon_{x^2-y^2} - \epsilon_{3z^2-r^2} \vert$, then Eq.~\eqref{eq:H_d_eff_3} can be reasonably approximated by Eq.~(3) of the main text.

	\section{Additional numerical results for T\MakeLowercase{i}A\MakeLowercase{u}$_4$}

	\subsection{Spin conductivity}
	
	As demonstrated in the main text, the orbital ($\bm{\sigma}^\mathbf{L}$) and spin ($\bm{\sigma}^\mathbf{S}$) conductivity tensors of TiAu$_4$ with  space group $I4/m$ take the following forms:
	\begin{gather}
		\bm{\sigma}^{X_x} =
		\begin{pmatrix}
			0 & 0 & \sigma_{xz}^{X_x} \\
			0 & 0 & \sigma_{yz}^{X_x} \\
			\sigma_{zx}^{X_x} & -\sigma_{zx}^{X_y} & 0 
		\end{pmatrix},
		\
		\bm{\sigma}^{X_y} =
		\begin{pmatrix}
			0 & 0 & -\sigma_{yz}^{X_x} \\
			0 & 0 & \sigma_{xz}^{X_x} \\
			\sigma_{zx}^{X_y} & \sigma_{zx}^{X_x} & 0 
		\end{pmatrix}, \
		\bm{\sigma}^{X_z} =
		\begin{pmatrix}
			\sigma_{xx}^{X_z} & - \sigma_{yx}^{X_z} & 0 \\
			\sigma_{yx}^{X_z} & \sigma_{xx}^{X_z} & 0 \\
			0 & 0 & \sigma_{zz}^{X_z}
		\end{pmatrix}, \label{eq:cond_tensor}
	\end{gather}
	with $X=L$ or $S$.	Although there are 13 nonzero elements in total, only 7 of them are independent: three conventional Hall components ($\sigma_{yz}^{X_x}$, $\sigma_{zx}^{X_y}$, and $\sigma_{yx}^{X_z}$) and four rotation-induced components, including two longitudinal components ($\sigma_{xx}^{X_z}$ and $\sigma_{zz}^{X_z}$) and two unconventional Hall components ($\sigma_{xz}^{X_x}$ and $\sigma_{zx}^{X_x}$). All these terms, for both orbital and spin conductivities, are even under time-reversal operation.

	While these components of the orbital conductivity tensor were presented in the main text (Fig.~4c and 4d), here we provide the corresponding components of the spin conductivity tensor, calculated using the Kubo formula:
	\begin{equation}\label{eq:kubo}
		\sigma_{\alpha \beta}^{X_\gamma} = 
		-e\hbar \int \frac{d^3\mathbf{k}}{(2\pi)^3} \sum_{n \neq m} 
		(f_{n\mathbf{k}} - f_{m\mathbf{k}}) \,  \mathrm{Im}
		\left[
		\frac
		{\langle \psi_{n\mathbf{k}} \vert \hat{J}_\alpha^{X_\gamma}  \vert \psi_ {m\mathbf{k}}\rangle 
			\langle \psi_{m\mathbf{k}} \vert \hat{v}_\beta \vert \psi_{n\mathbf{k}} \rangle 
		}
		{(\epsilon_{n\mathbf{k}} - \epsilon_{m\mathbf{k}})( \epsilon_{n\mathbf{k}} - \epsilon_{m\mathbf{k}} + i\Gamma )} 
		\right],
	\end{equation}
	where $\vert \psi_{n\mathbf{k}} \rangle$ is the eigenstate, $\epsilon_{n\mathbf{k}}$ is the energy eigenvalue, $f_{n\mathbf{k}} = 1/[e^{ (\epsilon_{n\mathbf{k}} - \mu ) / k_\mathrm{B}T} + 1 ]$ is the Fermi-Dirac distribution function, with the chemical potential $\mu$ and $k_\mathrm{B}T = 25$~meV, $\hat{J}_\alpha^{X_\gamma} = \frac{1}{2} (\hat{v}_\alpha \hat{X}_\gamma +  \hat{X}_\gamma \hat{v}_\alpha) $ is the conventional orbital ($X=L$) or spin ($X=S$) current operator, and $\hat{v}_\alpha $ is the $\alpha$-component of the velocity operator, evaluated by $\hat{v}_\alpha = \frac{1}{i\hbar} [\hat{r}_\alpha, \hat{\mathcal{H}}]$ with the position operator $\hat{r}_\alpha$. The lifetime broadening energy is set to $\Gamma =0.1$~eV.

\begin{figure}[h]
		\center\includegraphics[width=1.0\textwidth]{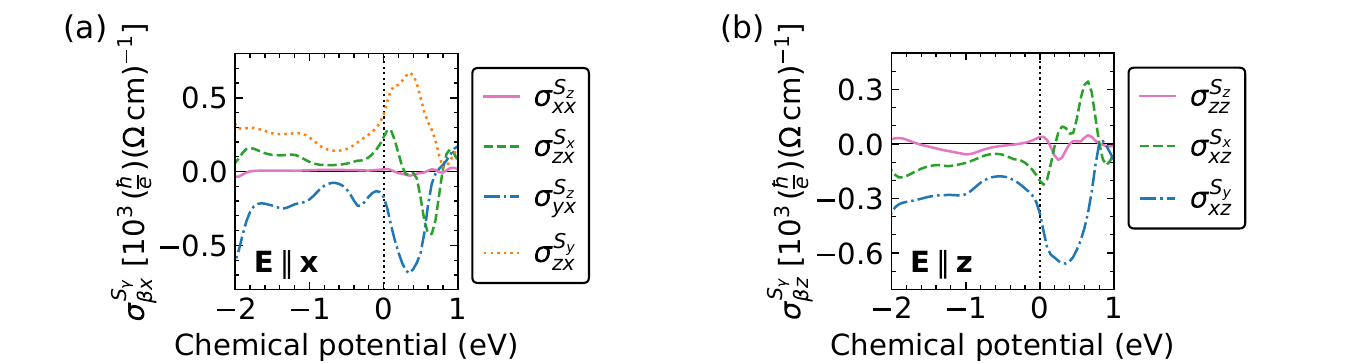}
		\caption{Nonzero spin conductivity components $\sigma_{\beta \alpha}^{S_\gamma}$ of TiAu$_4$ as functions of the chemical potential, for an electric field $\mathbf{E}$ applied along (a) the $x$ direction and (b) the $z$ direction.}
		\label{figS1} 
	\end{figure}

	Figures~\ref{figS1}(a) and \ref{figS1}(b) show the numerical results as functions of the chemical potential. The calculated values are as follows: for the conventional spin Hall components, $\sigma_{yx}^{S_z} = -170 \ (\hbar/e) (\Omega \, \mathrm{cm})^{-1}$, $\sigma_{zx}^{S_y} = 380 \ (\hbar/e) (\Omega \, \mathrm{cm})^{-1}$, and $\sigma_{xz}^{S_y} = -330 \ (\hbar/e) (\Omega \, \mathrm{cm})^{-1}$, for the longitudinal components, $\sigma_{xx}^{S_z} = 20 \ (\hbar/e) (\Omega \, \mathrm{cm})^{-1}$ and $\sigma_{zz}^{S_z} = 40 \ (\hbar/e) (\Omega \, \mathrm{cm})^{-1}$, and for the unconventional spin Hall components, $\sigma_{zx}^{S_x} = 230 \ (\hbar/e) (\Omega \, \mathrm{cm})^{-1}$ and $\sigma_{xz}^{S_x} = -200 \ (\hbar/e) (\Omega \, \mathrm{cm})^{-1}$. Note that all these components vanish in the absence of SOC, indicating that the spin currents in this system are generated through the relativistic effect. In particular, the rotation-induced components arise from the interaction between the spin and the FR order, which requires SOC, as demonstrated in Refs.~\cite{hayami2022electric, hayami2023unconventional}.

	\subsection{Effect of spin-orbit coupling on orbital conductivity}

\begin{figure}[h]
		\center\includegraphics[width=1.0\textwidth]{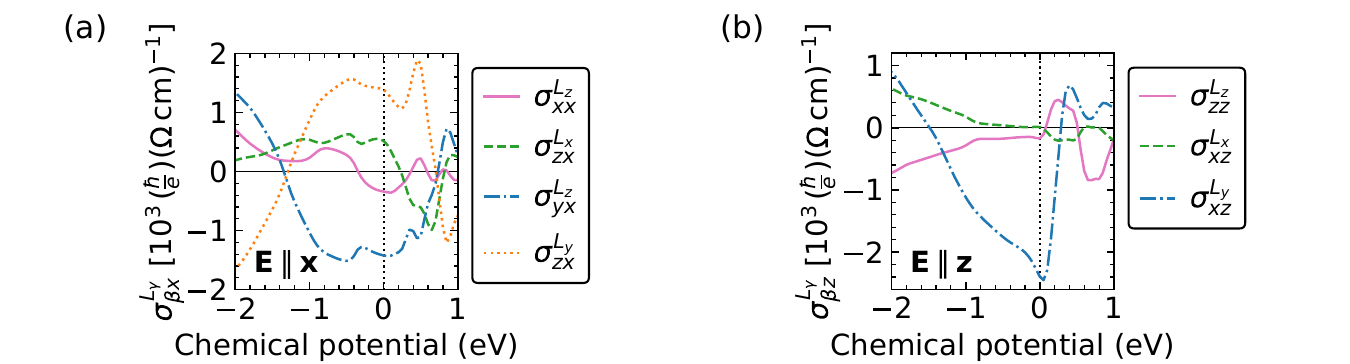}
		\caption{Nonzero orbital conductivity components $\sigma_{\beta \alpha}^{L_\gamma}$ of TiAu$_4$ without SOC, as functions of the chemical potential, for an electric field $\mathbf{E}$ applied along (a) the $x$ direction and (b) the $z$ direction.}
		\label{figS2} 
	\end{figure}

	In contrast to the spin conductivity, the orbital conductivity of TiAu$_4$ originates from a nonrelativistic mechanism. Therefore, the orbital conductivity is expected to be largely unaffected by the strength of SOC. To verify this, we calculated the orbital conductivity of TiAu$_4$ without SOC. Figures~\ref{figS2}(a) and \ref{figS2}(b) show the numerical results, which exhibit minimal deviation from those obtained with SOC (Fig.~4c and 4d of the main text). These results suggest that rotation-induced orbital currents can be observed across a broad range of materials, without the necessity for heavy elements with strong SOC. 
	
	\subsection{Dependence on lifetime broadening}
	
\begin{figure}[h]
		\center\includegraphics[width=1.0\textwidth]{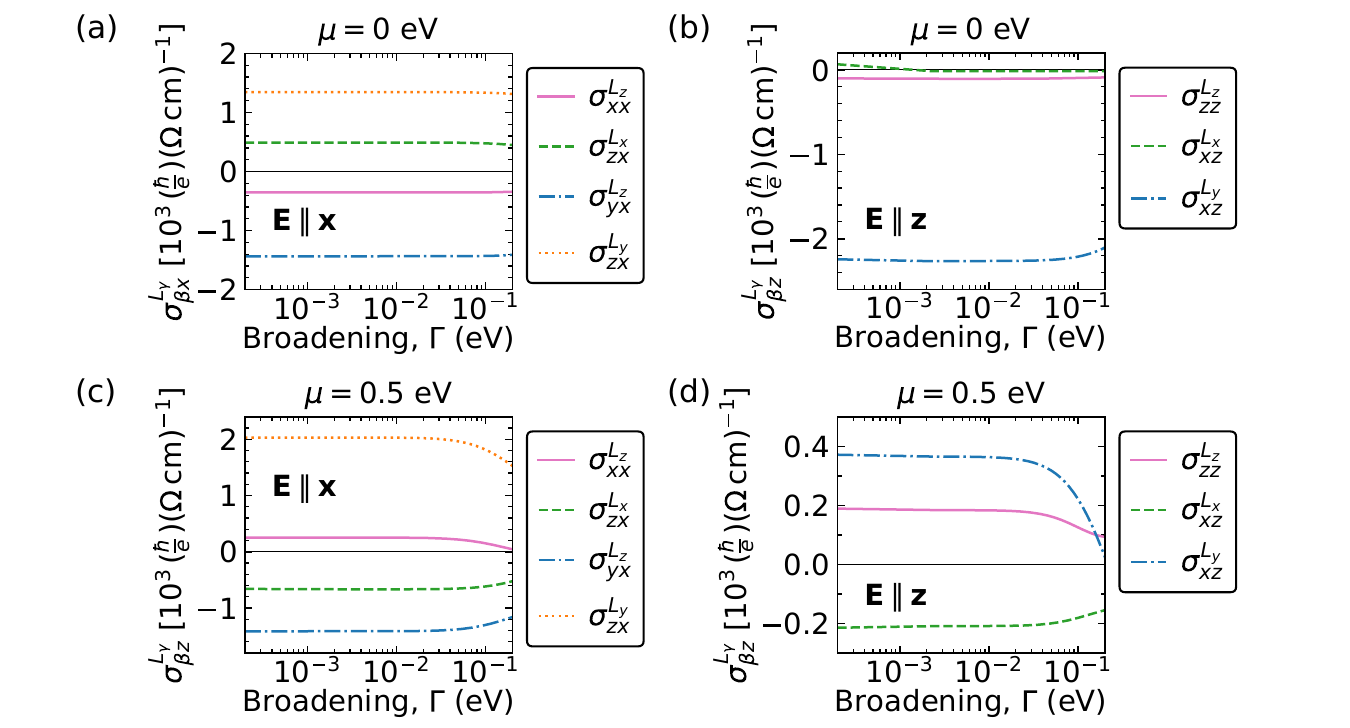}
		\caption{Nonzero orbital conductivity components $\sigma_{\beta \alpha}^{L_\gamma}$ of TiAu$_4$ as functions of the lifetime broadening $\Gamma$, for an electric field $\mathbf{E}$ applied along (a) the $x$ direction and (b) the $z$ direction.}
		\label{figS3} 
	\end{figure}

	The $\mathcal{T}$-even orbital conductivity in Eq.~\eqref{eq:kubo} arises from the interband contribution, indicating that the rotation-induced orbital currents originate from the intrinsic band structure rather than impurity scattering. To illustrate this, we investigated the dependence of the orbital conductivity on the lifetime broadening energy $\Gamma$ in Eq.~\eqref{eq:kubo}. Figures~\ref{figS3}(a) and \ref{figS3}(b) show that the $\mathcal{T}$-even orbital currents remain stable across varying $\Gamma$. Since $\sigma_{xz}^{L_x}$ is too small to clearly observe its dependence, we additionally calculated the orbital conductivity at the chemical potential of $\mu = 0.5$~eV. The results in Figs.~\ref{figS3}(c) and \ref{figS3}(d) also show the robustness of the $\mathcal{T}$-even orbital conductivity against $\Gamma $ up to the order of 0.1~eV. This behavior is in stark contrast to that of the $\mathcal{T}$-odd orbital or spin conductivity~\cite{salemi2022theory}, which exhibits a $1/\Gamma$ dependence at low $\Gamma$.

	\clearpage

	\section{Symmetry analysis for crystallographic point groups}
	
	Here we present the symmetry-allowed orbital (spin) conductivity tensors $\bm{\sigma}^\mathbf{L}$ ($\bm{\sigma}^\mathbf{S}$) for all 32 crystallographic point groups. The orbital (spin) conductivity is defined by 
	\begin{equation}\label{eq:linear}
		J_\alpha^{X_\gamma} = \sigma_{\alpha\beta}^{X_\gamma} E_\beta, \qquad (X=L,S)
	\end{equation}
	where $J_\alpha^{L_\gamma(S_\gamma)}$ is the orbital (spin) current flowing in the $\alpha$ direction and carrying orbital (spin) angular momentum polarized along the $\gamma$ direction, and $E_\beta$ is the electric field. Under a symmetry operation represented by an orthogonal matrix $\mathbf{R}$, $E_\beta$ transforms as
	\begin{equation}\label{eq:E_trans}
		\tilde{E}_\beta =  R_{\beta j} E_j,
	\end{equation}
	where Einstein summation is implied and the tilde denotes the quantities in the transformed coordinate system. Meanwhile, $J_\alpha^{X_\gamma}$ contains the angular momentum, thus it transforms as an axial tensor,
	\begin{equation}\label{eq:J_trans}
		\tilde{J}_\alpha^{X_\gamma}  = \det( \mathbf{R} ) R_{\alpha i} R_{\gamma k} J_i^{X_k}.
	\end{equation}
	To find how the orbital or spin conductivity tensor $\sigma_{\alpha\beta}^{X_\gamma}$ transforms, we rewrite the linear-response relation in Eq.~\eqref{eq:linear} in the transformed coordinates, 
	\begin{equation}\label{eq:linear_trans}
		\tilde{J}_\alpha^{X_\gamma} = \tilde{\sigma}_{\alpha\beta}^{X_\gamma} \tilde{E}_\beta .
	\end{equation}
	Using Eq.~\eqref{eq:J_trans}, $\tilde{J}_\alpha^{X_\gamma}$ can also be written as
	\begin{align} \label{eq:J_linear_trans}
		\tilde{J}_\alpha^{X_\gamma} & = \det( \mathbf{R} ) R_{\alpha i} R_{\gamma k} J_i^{X_k} \nonumber \\
		& =  \det( \mathbf{R} ) R_{\alpha i} R_{\gamma k}  \sigma_{i j}^{X_k} E_j \nonumber \\
		& = \det( \mathbf{R} ) R_{\alpha i} R_{\beta j} R_{\gamma k}  \sigma_{i j}^{X_k} \tilde{E}_\beta ,
	\end{align}
	where in the last line we used the inverse transformation of Eq.~\eqref{eq:E_trans}, $E_\beta =  R_{j \beta} \tilde{E}_j$. By comparing Eqs.~\eqref{eq:linear_trans} and \eqref{eq:J_linear_trans}, we obtain
	\begin{equation} \label{eq:sigma_trans}
		\tilde{\sigma}_{\alpha \beta}^{X_\gamma} = \det( \mathbf{R} ) R_{\alpha i} R_{\beta j} R_{\gamma k}  \sigma_{i j}^{X_k} .
	\end{equation}
	Neumann's principle states that a physical response tensor must be invariant under all symmetry operations of the crystal. Therefore, for every symmetry operation $\mathbf{R}$ belonging to the point group, the following condition must be satisfied:
	\begin{equation} \label{eq:neumann}
		\sigma_{\alpha \beta}^{X_\gamma} = \det( \mathbf{R} ) R_{\alpha i} R_{\beta j} R_{\gamma k}  \sigma_{i j}^{X_k}.
	\end{equation}
	We employ Eq.~\eqref{eq:neumann} to determine the symmetry constraints on $\sigma_{\alpha \beta}^{X_\gamma}$ for each crystallographic point group, using a set of generators for the group. The results are summarized in Table~\ref{tab:orbital_conductivity_tensors}.

	\hspace{2cm}

	\setlength{\LTcapwidth}{\textwidth}
	\newcommand{\tablespacing}{\rule{0pt}{8.7ex}}
	\begin{longtable}{
			c
			>{\centering\arraybackslash}p{0.26\textwidth}
			>{\centering\arraybackslash}p{0.26\textwidth}
			>{\centering\arraybackslash}p{0.26\textwidth}
		}
		\caption{%
			Symmetry-allowed orbital and spin conductivity tensors for all crystallographic point groups. The matrices are written in terms of a minimal set of independent components. 
		}
		\label{tab:orbital_conductivity_tensors}
		\\
		
		\hline
		Point group & $\bm{\sigma}^{X_x}$ & $\bm{\sigma}^{X_y}$ & $\bm{\sigma}^{X_z}$ \\
		\hline 
		\endfirsthead
		
		\hline
		Point group & $\bm{\sigma}^{X_x}$ & $\bm{\sigma}^{X_y}$ & $\bm{\sigma}^{X_z}$ \\
		\hline
		\endhead
		\tablespacing$1$ & $\begin{pmatrix}\sigma_{xx}^{X_x} & \sigma_{xy}^{X_x} & \sigma_{xz}^{X_x}\\\sigma_{yx}^{X_x} & \sigma_{yy}^{X_x} & \sigma_{yz}^{X_x}\\\sigma_{zx}^{X_x} & \sigma_{zy}^{X_x} & \sigma_{zz}^{X_x}\\\end{pmatrix}$ & $\begin{pmatrix}\sigma_{xx}^{X_y} & \sigma_{xy}^{X_y} & \sigma_{xz}^{X_y}\\\sigma_{yx}^{X_y} & \sigma_{yy}^{X_y} & \sigma_{yz}^{X_y}\\\sigma_{zx}^{X_y} & \sigma_{zy}^{X_y} & \sigma_{zz}^{X_y}\\\end{pmatrix}$ & $\begin{pmatrix}\sigma_{xx}^{X_z} & \sigma_{xy}^{X_z} & \sigma_{xz}^{X_z}\\\sigma_{yx}^{X_z} & \sigma_{yy}^{X_z} & \sigma_{yz}^{X_z}\\\sigma_{zx}^{X_z} & \sigma_{zy}^{X_z} & \sigma_{zz}^{X_z}\\\end{pmatrix}$ \\
		\tablespacing$\bar{1}$ & $\begin{pmatrix}\sigma_{xx}^{X_x} & \sigma_{xy}^{X_x} & \sigma_{xz}^{X_x}\\\sigma_{yx}^{X_x} & \sigma_{yy}^{X_x} & \sigma_{yz}^{X_x}\\\sigma_{zx}^{X_x} & \sigma_{zy}^{X_x} & \sigma_{zz}^{X_x}\\\end{pmatrix}$ & $\begin{pmatrix}\sigma_{xx}^{X_y} & \sigma_{xy}^{X_y} & \sigma_{xz}^{X_y}\\\sigma_{yx}^{X_y} & \sigma_{yy}^{X_y} & \sigma_{yz}^{X_y}\\\sigma_{zx}^{X_y} & \sigma_{zy}^{X_y} & \sigma_{zz}^{X_y}\\\end{pmatrix}$ & $\begin{pmatrix}\sigma_{xx}^{X_z} & \sigma_{xy}^{X_z} & \sigma_{xz}^{X_z}\\\sigma_{yx}^{X_z} & \sigma_{yy}^{X_z} & \sigma_{yz}^{X_z}\\\sigma_{zx}^{X_z} & \sigma_{zy}^{X_z} & \sigma_{zz}^{X_z}\\\end{pmatrix}$ \\
		\tablespacing$2$ & $\begin{pmatrix}0 & \sigma_{xy}^{X_x} & 0\\\sigma_{yx}^{X_x} & 0 & \sigma_{yz}^{X_x}\\0 & \sigma_{zy}^{X_x} & 0\\\end{pmatrix}$ & $\begin{pmatrix}\sigma_{xx}^{X_y} & 0 & \sigma_{xz}^{X_y}\\0 & \sigma_{yy}^{X_y} & 0\\\sigma_{zx}^{X_y} & 0 & \sigma_{zz}^{X_y}\\\end{pmatrix}$ & $\begin{pmatrix}0 & \sigma_{xy}^{X_z} & 0\\\sigma_{yx}^{X_z} & 0 & \sigma_{yz}^{X_z}\\0 & \sigma_{zy}^{X_z} & 0\\\end{pmatrix}$ \\
		\tablespacing$m$ & $\begin{pmatrix}0 & \sigma_{xy}^{X_x} & 0\\\sigma_{yx}^{X_x} & 0 & \sigma_{yz}^{X_x}\\0 & \sigma_{zy}^{X_x} & 0\\\end{pmatrix}$ & $\begin{pmatrix}\sigma_{xx}^{X_y} & 0 & \sigma_{xz}^{X_y}\\0 & \sigma_{yy}^{X_y} & 0\\\sigma_{zx}^{X_y} & 0 & \sigma_{zz}^{X_y}\\\end{pmatrix}$ & $\begin{pmatrix}0 & \sigma_{xy}^{X_z} & 0\\\sigma_{yx}^{X_z} & 0 & \sigma_{yz}^{X_z}\\0 & \sigma_{zy}^{X_z} & 0\\\end{pmatrix}$ \\
		\tablespacing$2/m$ & $\begin{pmatrix}0 & \sigma_{xy}^{X_x} & 0\\\sigma_{yx}^{X_x} & 0 & \sigma_{yz}^{X_x}\\0 & \sigma_{zy}^{X_x} & 0\\\end{pmatrix}$ & $\begin{pmatrix}\sigma_{xx}^{X_y} & 0 & \sigma_{xz}^{X_y}\\0 & \sigma_{yy}^{X_y} & 0\\\sigma_{zx}^{X_y} & 0 & \sigma_{zz}^{X_y}\\\end{pmatrix}$ & $\begin{pmatrix}0 & \sigma_{xy}^{X_z} & 0\\\sigma_{yx}^{X_z} & 0 & \sigma_{yz}^{X_z}\\0 & \sigma_{zy}^{X_z} & 0\\\end{pmatrix}$ \\
		\tablespacing$222$ & $\begin{pmatrix}0 & 0 & 0\\0 & 0 & \sigma_{yz}^{X_x}\\0 & \sigma_{zy}^{X_x} & 0\\\end{pmatrix}$ & $\begin{pmatrix}0 & 0 & \sigma_{xz}^{X_y}\\0 & 0 & 0\\\sigma_{zx}^{X_y} & 0 & 0\\\end{pmatrix}$ & $\begin{pmatrix}0 & \sigma_{xy}^{X_z} & 0\\\sigma_{yx}^{X_z} & 0 & 0\\0 & 0 & 0\\\end{pmatrix}$ \\
		\tablespacing$mm2$ & $\begin{pmatrix}0 & 0 & 0\\0 & 0 & \sigma_{yz}^{X_x}\\0 & \sigma_{zy}^{X_x} & 0\\\end{pmatrix}$ & $\begin{pmatrix}0 & 0 & \sigma_{xz}^{X_y}\\0 & 0 & 0\\\sigma_{zx}^{X_y} & 0 & 0\\\end{pmatrix}$ & $\begin{pmatrix}0 & \sigma_{xy}^{X_z} & 0\\\sigma_{yx}^{X_z} & 0 & 0\\0 & 0 & 0\\\end{pmatrix}$ \\
		\tablespacing$mmm$ & $\begin{pmatrix}0 & 0 & 0\\0 & 0 & \sigma_{yz}^{X_x}\\0 & \sigma_{zy}^{X_x} & 0\\\end{pmatrix}$ & $\begin{pmatrix}0 & 0 & \sigma_{xz}^{X_y}\\0 & 0 & 0\\\sigma_{zx}^{X_y} & 0 & 0\\\end{pmatrix}$ & $\begin{pmatrix}0 & \sigma_{xy}^{X_z} & 0\\\sigma_{yx}^{X_z} & 0 & 0\\0 & 0 & 0\\\end{pmatrix}$ \\
		\tablespacing$4$ & $\begin{pmatrix}0 & 0 & \sigma_{xz}^{X_x}\\0 & 0 & -\sigma_{xz}^{X_y}\\\sigma_{zx}^{X_x} & -\sigma_{zx}^{X_y} & 0\\\end{pmatrix}$ & $\begin{pmatrix}0 & 0 & \sigma_{xz}^{X_y}\\0 & 0 & \sigma_{xz}^{X_x}\\\sigma_{zx}^{X_y} & \sigma_{zx}^{X_x} & 0\\\end{pmatrix}$ & $\begin{pmatrix}\sigma_{xx}^{X_z} & \sigma_{xy}^{X_z} & 0\\-\sigma_{xy}^{X_z} & \sigma_{xx}^{X_z} & 0\\0 & 0 & \sigma_{zz}^{X_z}\\\end{pmatrix}$ \\
		\tablespacing$\bar{4}$ & $\begin{pmatrix}0 & 0 & \sigma_{xz}^{X_x}\\0 & 0 & -\sigma_{xz}^{X_y}\\\sigma_{zx}^{X_x} & -\sigma_{zx}^{X_y} & 0\\\end{pmatrix}$ & $\begin{pmatrix}0 & 0 & \sigma_{xz}^{X_y}\\0 & 0 & \sigma_{xz}^{X_x}\\\sigma_{zx}^{X_y} & \sigma_{zx}^{X_x} & 0\\\end{pmatrix}$ & $\begin{pmatrix}\sigma_{xx}^{X_z} & \sigma_{xy}^{X_z} & 0\\-\sigma_{xy}^{X_z} & \sigma_{xx}^{X_z} & 0\\0 & 0 & \sigma_{zz}^{X_z}\\\end{pmatrix}$ \\
		\tablespacing$4/m$ & $\begin{pmatrix}0 & 0 & \sigma_{xz}^{X_x}\\0 & 0 & -\sigma_{xz}^{X_y}\\\sigma_{zx}^{X_x} & -\sigma_{zx}^{X_y} & 0\\\end{pmatrix}$ & $\begin{pmatrix}0 & 0 & \sigma_{xz}^{X_y}\\0 & 0 & \sigma_{xz}^{X_x}\\\sigma_{zx}^{X_y} & \sigma_{zx}^{X_x} & 0\\\end{pmatrix}$ & $\begin{pmatrix}\sigma_{xx}^{X_z} & \sigma_{xy}^{X_z} & 0\\-\sigma_{xy}^{X_z} & \sigma_{xx}^{X_z} & 0\\0 & 0 & \sigma_{zz}^{X_z}\\\end{pmatrix}$ \\
		\tablespacing$422$ & $\begin{pmatrix}0 & 0 & 0\\0 & 0 & -\sigma_{xz}^{X_y}\\0 & -\sigma_{zx}^{X_y} & 0\\\end{pmatrix}$ & $\begin{pmatrix}0 & 0 & \sigma_{xz}^{X_y}\\0 & 0 & 0\\\sigma_{zx}^{X_y} & 0 & 0\\\end{pmatrix}$ & $\begin{pmatrix}0 & \sigma_{xy}^{X_z} & 0\\-\sigma_{xy}^{X_z} & 0 & 0\\0 & 0 & 0\\\end{pmatrix}$ \\
		\tablespacing$4mm$ & $\begin{pmatrix}0 & 0 & 0\\0 & 0 & -\sigma_{xz}^{X_y}\\0 & -\sigma_{zx}^{X_y} & 0\\\end{pmatrix}$ & $\begin{pmatrix}0 & 0 & \sigma_{xz}^{X_y}\\0 & 0 & 0\\\sigma_{zx}^{X_y} & 0 & 0\\\end{pmatrix}$ & $\begin{pmatrix}0 & \sigma_{xy}^{X_z} & 0\\-\sigma_{xy}^{X_z} & 0 & 0\\0 & 0 & 0\\\end{pmatrix}$ \\
		\tablespacing$\bar{4}$2m & $\begin{pmatrix}0 & 0 & 0\\0 & 0 & -\sigma_{xz}^{X_y}\\0 & -\sigma_{zx}^{X_y} & 0\\\end{pmatrix}$ & $\begin{pmatrix}0 & 0 & \sigma_{xz}^{X_y}\\0 & 0 & 0\\\sigma_{zx}^{X_y} & 0 & 0\\\end{pmatrix}$ & $\begin{pmatrix}0 & \sigma_{xy}^{X_z} & 0\\-\sigma_{xy}^{X_z} & 0 & 0\\0 & 0 & 0\\\end{pmatrix}$ \\
		\tablespacing$4/mmm$ & $\begin{pmatrix}0 & 0 & 0\\0 & 0 & -\sigma_{xz}^{X_y}\\0 & -\sigma_{zx}^{X_y} & 0\\\end{pmatrix}$ & $\begin{pmatrix}0 & 0 & \sigma_{xz}^{X_y}\\0 & 0 & 0\\\sigma_{zx}^{X_y} & 0 & 0\\\end{pmatrix}$ & $\begin{pmatrix}0 & \sigma_{xy}^{X_z} & 0\\-\sigma_{xy}^{X_z} & 0 & 0\\0 & 0 & 0\\\end{pmatrix}$ \\
		\tablespacing$3$ & $\begin{pmatrix}\sigma_{xx}^{X_x} & \sigma_{xx}^{X_y} & \sigma_{xz}^{X_x}\\\sigma_{xx}^{X_y} & -\sigma_{xx}^{X_x} & -\sigma_{xz}^{X_y}\\\sigma_{zx}^{X_x} & -\sigma_{zx}^{X_y} & 0\\\end{pmatrix}$ & $\begin{pmatrix}\sigma_{xx}^{X_y} & -\sigma_{xx}^{X_x} & \sigma_{xz}^{X_y}\\-\sigma_{xx}^{X_x} & -\sigma_{xx}^{X_y} & \sigma_{xz}^{X_x}\\\sigma_{zx}^{X_y} & \sigma_{zx}^{X_x} & 0\\\end{pmatrix}$ & $\begin{pmatrix}\sigma_{xx}^{X_z} & \sigma_{xy}^{X_z} & 0\\-\sigma_{xy}^{X_z} & \sigma_{xx}^{X_z} & 0\\0 & 0 & \sigma_{zz}^{X_z}\\\end{pmatrix}$ \\
		\tablespacing$\bar{3}$ & $\begin{pmatrix}\sigma_{xx}^{X_x} & \sigma_{xx}^{X_y} & \sigma_{xz}^{X_x}\\\sigma_{xx}^{X_y} & -\sigma_{xx}^{X_x} & -\sigma_{xz}^{X_y}\\\sigma_{zx}^{X_x} & -\sigma_{zx}^{X_y} & 0\\\end{pmatrix}$ & $\begin{pmatrix}\sigma_{xx}^{X_y} & -\sigma_{xx}^{X_x} & \sigma_{xz}^{X_y}\\-\sigma_{xx}^{X_x} & -\sigma_{xx}^{X_y} & \sigma_{xz}^{X_x}\\\sigma_{zx}^{X_y} & \sigma_{zx}^{X_x} & 0\\\end{pmatrix}$ & $\begin{pmatrix}\sigma_{xx}^{X_z} & \sigma_{xy}^{X_z} & 0\\-\sigma_{xy}^{X_z} & \sigma_{xx}^{X_z} & 0\\0 & 0 & \sigma_{zz}^{X_z}\\\end{pmatrix}$ \\
		\tablespacing32 & $\begin{pmatrix}\sigma_{xx}^{X_x} & 0 & 0\\0 & -\sigma_{xx}^{X_x} & -\sigma_{xz}^{X_y}\\0 & -\sigma_{zx}^{X_y} & 0\\\end{pmatrix}$ & $\begin{pmatrix}0 & -\sigma_{xx}^{X_x} & \sigma_{xz}^{X_y}\\-\sigma_{xx}^{X_x} & 0 & 0\\\sigma_{zx}^{X_y} & 0 & 0\\\end{pmatrix}$ & $\begin{pmatrix}0 & \sigma_{xy}^{X_z} & 0\\-\sigma_{xy}^{X_z} & 0 & 0\\0 & 0 & 0\\\end{pmatrix}$ \\
		\tablespacing$3m$ & $\begin{pmatrix}\sigma_{xx}^{X_x} & 0 & 0\\0 & -\sigma_{xx}^{X_x} & -\sigma_{xz}^{X_y}\\0 & -\sigma_{zx}^{X_y} & 0\\\end{pmatrix}$ & $\begin{pmatrix}0 & -\sigma_{xx}^{X_x} & \sigma_{xz}^{X_y}\\-\sigma_{xx}^{X_x} & 0 & 0\\\sigma_{zx}^{X_y} & 0 & 0\\\end{pmatrix}$ & $\begin{pmatrix}0 & \sigma_{xy}^{X_z} & 0\\-\sigma_{xy}^{X_z} & 0 & 0\\0 & 0 & 0\\\end{pmatrix}$ \\
		\tablespacing$\bar{3}m$ & $\begin{pmatrix}\sigma_{xx}^{X_x} & 0 & 0\\0 & -\sigma_{xx}^{X_x} & -\sigma_{xz}^{X_y}\\0 & -\sigma_{zx}^{X_y} & 0\\\end{pmatrix}$ & $\begin{pmatrix}0 & -\sigma_{xx}^{X_x} & \sigma_{xz}^{X_y}\\-\sigma_{xx}^{X_x} & 0 & 0\\\sigma_{zx}^{X_y} & 0 & 0\\\end{pmatrix}$ & $\begin{pmatrix}0 & \sigma_{xy}^{X_z} & 0\\-\sigma_{xy}^{X_z} & 0 & 0\\0 & 0 & 0\\\end{pmatrix}$ \\
		\tablespacing$6$ & $\begin{pmatrix}0 & 0 & \sigma_{xz}^{X_x}\\0 & 0 & -\sigma_{xz}^{X_y}\\\sigma_{zx}^{X_x} & -\sigma_{zx}^{X_y} & 0\\\end{pmatrix}$ & $\begin{pmatrix}0 & 0 & \sigma_{xz}^{X_y}\\0 & 0 & \sigma_{xz}^{X_x}\\\sigma_{zx}^{X_y} & \sigma_{zx}^{X_x} & 0\\\end{pmatrix}$ & $\begin{pmatrix}\sigma_{xx}^{X_z} & \sigma_{xy}^{X_z} & 0\\-\sigma_{xy}^{X_z} & \sigma_{xx}^{X_z} & 0\\0 & 0 & \sigma_{zz}^{X_z}\\\end{pmatrix}$ \\
		\tablespacing$\bar{6}$ & $\begin{pmatrix}0 & 0 & \sigma_{xz}^{X_x}\\0 & 0 & -\sigma_{xz}^{X_y}\\\sigma_{zx}^{X_x} & -\sigma_{zx}^{X_y} & 0\\\end{pmatrix}$ & $\begin{pmatrix}0 & 0 & \sigma_{xz}^{X_y}\\0 & 0 & \sigma_{xz}^{X_x}\\\sigma_{zx}^{X_y} & \sigma_{zx}^{X_x} & 0\\\end{pmatrix}$ & $\begin{pmatrix}\sigma_{xx}^{X_z} & \sigma_{xy}^{X_z} & 0\\-\sigma_{xy}^{X_z} & \sigma_{xx}^{X_z} & 0\\0 & 0 & \sigma_{zz}^{X_z}\\\end{pmatrix}$ \\
		\tablespacing$6/m$ & $\begin{pmatrix}0 & 0 & \sigma_{xz}^{X_x}\\0 & 0 & -\sigma_{xz}^{X_y}\\\sigma_{zx}^{X_x} & -\sigma_{zx}^{X_y} & 0\\\end{pmatrix}$ & $\begin{pmatrix}0 & 0 & \sigma_{xz}^{X_y}\\0 & 0 & \sigma_{xz}^{X_x}\\\sigma_{zx}^{X_y} & \sigma_{zx}^{X_x} & 0\\\end{pmatrix}$ & $\begin{pmatrix}\sigma_{xx}^{X_z} & \sigma_{xy}^{X_z} & 0\\-\sigma_{xy}^{X_z} & \sigma_{xx}^{X_z} & 0\\0 & 0 & \sigma_{zz}^{X_z}\\\end{pmatrix}$ \\
		\tablespacing$622$ & $\begin{pmatrix}0 & 0 & 0\\0 & 0 & -\sigma_{xz}^{X_y}\\0 & -\sigma_{zx}^{X_y} & 0\\\end{pmatrix}$ & $\begin{pmatrix}0 & 0 & \sigma_{xz}^{X_y}\\0 & 0 & 0\\\sigma_{zx}^{X_y} & 0 & 0\\\end{pmatrix}$ & $\begin{pmatrix}0 & \sigma_{xy}^{X_z} & 0\\-\sigma_{xy}^{X_z} & 0 & 0\\0 & 0 & 0\\\end{pmatrix}$ \\
		\tablespacing$6mm$ & $\begin{pmatrix}0 & 0 & 0\\0 & 0 & -\sigma_{xz}^{X_y}\\0 & -\sigma_{zx}^{X_y} & 0\\\end{pmatrix}$ & $\begin{pmatrix}0 & 0 & \sigma_{xz}^{X_y}\\0 & 0 & 0\\\sigma_{zx}^{X_y} & 0 & 0\\\end{pmatrix}$ & $\begin{pmatrix}0 & \sigma_{xy}^{X_z} & 0\\-\sigma_{xy}^{X_z} & 0 & 0\\0 & 0 & 0\\\end{pmatrix}$ \\
		\tablespacing$\bar{6}$m2 & $\begin{pmatrix}0 & 0 & 0\\0 & 0 & -\sigma_{xz}^{X_y}\\0 & -\sigma_{zx}^{X_y} & 0\\\end{pmatrix}$ & $\begin{pmatrix}0 & 0 & \sigma_{xz}^{X_y}\\0 & 0 & 0\\\sigma_{zx}^{X_y} & 0 & 0\\\end{pmatrix}$ & $\begin{pmatrix}0 & \sigma_{xy}^{X_z} & 0\\-\sigma_{xy}^{X_z} & 0 & 0\\0 & 0 & 0\\\end{pmatrix}$ \\
		\tablespacing$6/mmm$ & $\begin{pmatrix}0 & 0 & 0\\0 & 0 & -\sigma_{xz}^{X_y}\\0 & -\sigma_{zx}^{X_y} & 0\\\end{pmatrix}$ & $\begin{pmatrix}0 & 0 & \sigma_{xz}^{X_y}\\0 & 0 & 0\\\sigma_{zx}^{X_y} & 0 & 0\\\end{pmatrix}$ & $\begin{pmatrix}0 & \sigma_{xy}^{X_z} & 0\\-\sigma_{xy}^{X_z} & 0 & 0\\0 & 0 & 0\\\end{pmatrix}$ \\
		\tablespacing$23$ & $\begin{pmatrix}0 & 0 & 0\\0 & 0 & \sigma_{xy}^{X_z}\\0 & \sigma_{xz}^{X_y} & 0\\\end{pmatrix}$ & $\begin{pmatrix}0 & 0 & \sigma_{xz}^{X_y}\\0 & 0 & 0\\\sigma_{xy}^{X_z} & 0 & 0\\\end{pmatrix}$ & $\begin{pmatrix}0 & \sigma_{xy}^{X_z} & 0\\\sigma_{xz}^{X_y} & 0 & 0\\0 & 0 & 0\\\end{pmatrix}$ \\
		\tablespacing$m\bar{3}$ & $\begin{pmatrix}0 & 0 & 0\\0 & 0 & \sigma_{xy}^{X_z}\\0 & \sigma_{xz}^{X_y} & 0\\\end{pmatrix}$ & $\begin{pmatrix}0 & 0 & \sigma_{xz}^{X_y}\\0 & 0 & 0\\\sigma_{xy}^{X_z} & 0 & 0\\\end{pmatrix}$ & $\begin{pmatrix}0 & \sigma_{xy}^{X_z} & 0\\\sigma_{xz}^{X_y} & 0 & 0\\0 & 0 & 0\\\end{pmatrix}$ \\
		\tablespacing$432$ & $\begin{pmatrix}0 & 0 & 0\\0 & 0 & \sigma_{xy}^{X_z}\\0 & -\sigma_{xy}^{X_z} & 0\\\end{pmatrix}$ & $\begin{pmatrix}0 & 0 & -\sigma_{xy}^{X_z}\\0 & 0 & 0\\\sigma_{xy}^{X_z} & 0 & 0\\\end{pmatrix}$ & $\begin{pmatrix}0 & \sigma_{xy}^{X_z} & 0\\-\sigma_{xy}^{X_z} & 0 & 0\\0 & 0 & 0\\\end{pmatrix}$ \\
		\tablespacing$\bar{4}3m$ & $\begin{pmatrix}0 & 0 & 0\\0 & 0 & \sigma_{xy}^{X_z}\\0 & -\sigma_{xy}^{X_z} & 0\\\end{pmatrix}$ & $\begin{pmatrix}0 & 0 & -\sigma_{xy}^{X_z}\\0 & 0 & 0\\\sigma_{xy}^{X_z} & 0 & 0\\\end{pmatrix}$ & $\begin{pmatrix}0 & \sigma_{xy}^{X_z} & 0\\-\sigma_{xy}^{X_z} & 0 & 0\\0 & 0 & 0\\\end{pmatrix}$ \\
		\tablespacing$m\bar{3}$m & $\begin{pmatrix}0 & 0 & 0\\0 & 0 & \sigma_{xy}^{X_z}\\0 & -\sigma_{xy}^{X_z} & 0\\\end{pmatrix}$ & $\begin{pmatrix}0 & 0 & -\sigma_{xy}^{X_z}\\0 & 0 & 0\\\sigma_{xy}^{X_z} & 0 & 0\\\end{pmatrix}$ & $\begin{pmatrix}0 & \sigma_{xy}^{X_z} & 0\\-\sigma_{xy}^{X_z} & 0 & 0\\0 & 0 & 0\\\end{pmatrix}$ \\
	\end{longtable}

	\clearpage

	\bibliography{ref.bib}